\documentclass[aps,prd,onecolumn,showpacs,showkeys,amsmath,amssymb]{revtex4}
\usepackage{graphicx,color}
\usepackage{bm}
\begin{document}
\vspace*{1.0cm}
\begin{flushright}
PKU-HEP\;\;08/2006\\
August 2006\\
arXiv: hep-ph/0609013
\end{flushright}
\title{\sf\Large{$D_{sJ}(2860)$ and $D_{sJ}(2715)$}}

\baselineskip 20pt
\author{Bo Zhang}
\author{Xiang Liu}
\email{liuxiang726@mail.nankai.edu.cn}
\author{Wei-Zhen Deng}
\author{Shi-Lin Zhu}
\email{zhusl@th.phy.pku.edu.cn}

\vspace{2.0cm}

\affiliation{Department of Physics, Peking University, Beijing
100871, China}

\vspace*{1.0cm}

\date{\today}
\begin{abstract}

Recently Babar Collaboration reported a new $c\bar{s}$ state
$D_{sJ}(2860)$ and Belle Collaboration observed $D_{sJ}(2715)$. We
investigate the strong decays of the excited $c\bar{s}$ states
using the $^{3}P_{0}$ model. After comparing the theoretical decay
widths and decay patterns with the available experimental data, we
tend to conclude: (1) $D_{sJ}(2715)$ is probably the
$1^{-}(1^{3}D_{1})$ $c\bar{s}$ state although the
$1^{-}(2^{3}S_{1})$ assignment is not completely excluded; (2)
$D_{sJ}(2860)$ seems unlikely to be the $1^{-}(2^{3}S_{1})$ and
$1^{-}(1^{3}D_{1})$ candidate; (3) $D_{sJ}(2860)$ as either a
$0^{+}(2^{3}P_{0})$ or $3^{-}(1^{3}D_{3})$ $c\bar{s}$ state is
consistent with the experimental data; (4) experimental search of
$D_{sJ}(2860)$ in the channels $D_s\eta$, $DK^{*}$, $D^{*}K$ and
$D_{s}^{*}\eta$ will be crucial to distinguish the above two
possibilities.

\end{abstract}



\maketitle

\newpage

\section{introduction}\label{sec1}

Recently, Babar collaboration observed a new $c\bar s$ state
$D_{sJ}(2860)$ with a mass $2856.6\pm1.5\pm 5.0$ MeV and width
$\Gamma=(48\pm 7\pm10)$ MeV. Babar observed it only in the
$D^{0}K^{+},D^{+}K_{S}^{0}$ channels and found no evidence
$D^{\ast 0}K^{+}$ and $D^{*+}K_{S}^{0}$. Thus its $J^{P}=0^{+},
1^{-}, 2^{+}, 3^{-}, \cdots$. At the same time, Belle
collaboration reported a broader $c\bar{s}$ state $D_{sJ}(2715)$
with $J^{P}=1^{-}$ in $B^{+}\to \bar{D}^{0}D^{0}K^{+}$ decay
\cite{belle-2715}. Its mass is $2715\pm 11^{+11}_{-14}$ MeV and
width $\Gamma=(115\pm 20^{+36}_{-32})$ MeV.

$D_{sJ}(2860)$ was proposed as the first radial excitation of
$D_{sJ}^{*}(2317)$ in Ref. \cite{beveren}, as a $J^{P}=3^{-}$
$c\bar{s}$ state in Ref. \cite{colangelo} and as $c\bar{s}(2P)$
state in Ref. \cite{close-2860}. $D_{sJ}(2715)$ sits exactly on
the quark model prediction $2720$ MeV for the $2^{3}S_{1}$ $c\bar
s$ state \cite{potential model}. The $1^-$ state lies around
$2721$ MeV if one requires the $(1^+, 1^-$) $c\bar s$ states form
a chiral doublet \cite{chiral}.

According to the heavy quark effective field theory, heavy mesons
form doublets. For example, we have one s-wave $c\bar s$ doublet
$(0^-, 1^-)=(D_{s}(1965),D_{s}^{*}(2115))$ and two p-wave doublets
$(0^+, 1^+)=(D_{sJ}^{*}(2317),D_{sJ}(2460))$ and $(1^+,
2^+)=(D_{s1}(2536),D_{s2}(2573))$. The two d-wave $c\bar s$
doublets $(1^-, 2^-)$ and $(2^-, 3^-)$ have not been observed yet.

The possible quantum numbers of $D_{sJ}(2860)$ include
$0^{+}(2^{3}P_{0})$, $1^{-} (1^{3}D_{1})$, $1^- (2 ^3S_1)$, $2^+
(2 ^3P_2)$, $2^+ (1 ^3F_2)$ and $3^{-}(1^{3}D_{3})$. The $2 ^3P_2$
$c\bar s$ state is expected to lie around $(2.95\sim 3.0)$ GeV
while the mass of the $1 ^3F_2$ state will be much higher than
2.86 GeV. In the following we focus on the $0^+, 1^-, 3^-$
assignments.

In this work, we will try different assignments for $D_{sJ}(2860)$
and $D_{sJ}(2715)$ and investigate their different decay patterns
and total widths within the framework of the $^{3}P_{0}$ model.
After comparing the theoretical total width and decay patterns
with the available experimental information, one may get a hint of
their favorable quantum numbers and assignments in the quark
model.

This paper is organized as follows. We give a brief review of
$^{3}P_{0}$ model in Section \ref{sec2}. Then we present the
strong decay amplitudes of various low-lying excited $c\bar s$
states in Sections \ref{sec3}-\ref{sec6}. We present our numerical
results in Section \ref{sec7}. The last section is the discussion.
We collect some lengthy formulae in the Appendix.

\section{The $^{3}P_{0}$ Model}\label{sec2}

The $^{3}P_{0}$ model was first proposed by Micu \cite{Micu} and
further developed by Yaouanc et al. later
\cite{yaouanc,yaouanc-1,yaouanc-book}. Now this model has been
widely used to study the hadron decay widths
\cite{qpc-1,qpc-2,qpc-90,ackleh,Zou,liu,lujie}.

\begin{figure}
\begin{center}
\scalebox{0.5}{\includegraphics{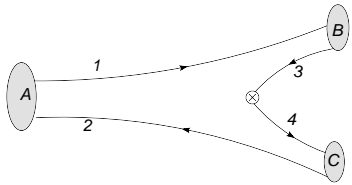}}
\end{center}
\caption{The decay process $A \rightarrow B C$ in the $^3P_0$
model. }\label{fig1}
\end{figure}

According to this model, a pair of quarks with $J^{PC}=0^{++}$ is
created from vacuum when a hadron decays, which is shown in Fig.
\ref{fig1} for the meson decay process $A\to BC$. The new
$q\bar{q}$ pair created from the vacuum together with the
$q\bar{q}$ within the the initial meson regroups into the outgoing
mesons via the quark rearrangement process. In the nonrelativistic
limit, the transition operator is written as
\begin{equation}
T = - 3 \gamma \sum_m\: \langle 1\;m;1\;-m|0\;0 \rangle\,
(2\pi)^{3\over2}\int\!{\rm d}^3{\textbf{k}}_3\; {\rm
d}^3{\textbf{k}}_4 \delta^3({\textbf{k}}_3+{\textbf{k}}_4)\: {\cal
Y}^m_1({\textbf{k}}_3-{\textbf{k}_4})\; \chi^{3 4}_{1, -\!m}\;
\varphi^{3 4}_0\;\, \omega^{3 4}_0\;
b^\dagger_{3i}({\textbf{k}}_3)\; d^\dagger_{4j}({\textbf{k}}_4)
\label{tmatrix}
\end{equation}
where $i$ and $j$ are the SU(3)-color indices of the created quark
and anti-quark. $\varphi^{34}_{0}=(u\bar u +d\bar d +s \bar
s)/\sqrt 3$ and $\omega_{0}^{34}=\delta_{ij}$ for flavor and color
singlets respectively. $\chi_{{1,-m}}^{34}$ is a triplet state of
spin. $\mathcal{Y}_{1}^{m}(\mathbf{k})\equiv
|\mathbf{k}|^{l}Y_{l}^{m}(\theta_{k},\phi_{k})$ is a solid
harmonic polynomial corresponding to the p-wave quark pair.
$\gamma$ is a dimensionless constant which denotes the strength of
quark pair creation from vacuum and can be extracted by fitting
data. The meson state is defined as \cite{mockmeson}
\begin{eqnarray}\label{mockmeson}
|A(n_A \mbox{}^{2S_A+1}L_A \,\mbox{}_{J_A M_{J_A}}) ({\textbf{p}}_A)
\rangle &=& \sqrt{2 E_A}\: \!\!\!\!\!\!\sum_{M_{L_A},M_{S_A}}\!\!\!
\langle L_A M_{L_A}
S_A M_{S_A} | J_A M_{J_A} \rangle \nonumber\\
&&\!\!\!\!\!\!\!\!\!\!\!\!\times \;\!\!\int\!{\rm
d}^3{\textbf{k}}_1{\rm
d}^3{\textbf{k}}_2\delta^3({\bf{k_1+k_2\!-\!p_A}})\psi_{n_A L_A
M_{L_A}}\!({\bf{k_1,k_2}})\chi^{1 2}_{S_A M_{S_A}}\varphi^{1
2}_A\omega^{1 2}_A |\;q_1({\textbf{k}}_1)
\bar{q}_2({\textbf{k}}_2)\rangle
\end{eqnarray}
and satisfies the normalization condition
\begin{eqnarray}
\langle A({\textbf{p}}_A)|A({\textbf{p}}'_A) \rangle &=& 2E_A
\delta^3({\textbf{p}}_A-{\textbf{p}}'_A)\; .
\end{eqnarray}
The subscripts 1 and 2 refer to the quark and antiquark within the
parent meson A respectively. ${\textbf{k}}_1$ and ${\textbf{k}}_2$
are the momentum of the quark and antiquark. ${\textbf{k}}_A$ is
their relative momentum and ${\textbf{p}}_A$ is the momentum of
meson A. $S_A=s_{q_1}+s_{q_2}$ is the total spin. $J_A=L_A+S_A$ is
the total angular momentum.

The S-matrix is defined as
\begin{eqnarray}
\langle
f|S|i\rangle&=&I+i(2\pi)^4\delta^4(p_f-p_i)\mathcal{M}^{M_{J_A}
M_{J_B} M_{J_C}}\;.
\end{eqnarray}
The decay helicity amplitude of the process $A\rightarrow B+C$ in
the meson A center of mass frame is
\begin{eqnarray}
&&{\mathcal{M}}^{M_{J_A} M_{J_B} M_{J_C}}(A\rightarrow BC)
\nonumber\\ &=&\sqrt{8 E_A E_B E_C}\;\;\gamma\!\!\!\!\!\!\!\!\!\!\!
\sum_{\renewcommand{\arraystretch}{.5}\begin{array}[t]{l}
\scriptstyle M_{L_A},M_{S_A},\\
\scriptstyle M_{L_B},M_{S_B},\\
\scriptstyle M_{L_C},M_{S_C},m
\end{array}}\renewcommand{\arraystretch}{1}\!\!\!\!\!\!\!\!
\langle L_A M_{L_A} S_A M_{S_A} | J_A M_{J_A} \rangle  \langle L_B
M_{L_B} S_B M_{S_B} | J_B M_{J_B} \rangle \langle L_C M_{L_C} S_C
M_{S_C} | J_C M_{J_C} \rangle \nonumber\\
&& \times  \langle 1\;m;1\;-m|\;0\;0 \rangle\; \langle \chi^{3
2}_{S_C M_{S_C}}\chi^{1 4}_{S_B M_{S_B}}  | \chi^{1 2}_{S_A M_{S_A}}
\chi^{3 4}_{1 -\!m} \rangle
 \langle\varphi^{3 2}_C \varphi^{1 4}_B | \varphi^{1 2}_A \varphi^{3 4}_0
 \rangle
\;I^{M_{L_A},m}_{M_{L_B},M_{L_C}}({\textbf{p}}) \;,\nonumber\\
\end{eqnarray}
where the spatial integral $I_m(A,BC)$ is defined as
\begin{eqnarray}
I^{M_{L_A},m}_{M_{L_B},M_{L_C}}({\textbf{p}})&=&\;{(2\pi)^{3/2}}
\int\!{\rm d}^3{\textbf{k}}_1{\rm d}^3{\textbf{k}}_2{\rm
d}^3{\textbf{k}}_3{\rm d}^3{\textbf{k}}_4\;
\delta^3({\bf{k_3+k_4}})\delta^3({\bf{k_1+k_2-p_{_A}}})\delta^3({\bf{k_1+k_3-p_{_B}}})\delta^3({\bf{k_2+k_4-p_{_C}}})
\nonumber\\
&&\times \psi^*_{n_B L_B M_{L_B}}\!
({\textbf{k}}_1,{\textbf{k}}_3)\; \psi^*_{n_C L_C M_{L_C}}\!
({\textbf{k}}_2,{\textbf{k}}_4)\; \psi_{n_A L_A M_{L_A}}\!
({\textbf{k}}_1,{\textbf{k}}_2)\; {\cal
Y}^m_1\big(\frac{{\bf{k_3-k_4}}}{2}\big). \label{integral}
\end{eqnarray}

The spin matrix element can be written in terms of Wigner's 9j
symbol \cite{yaouanc-book}
\begin{eqnarray}
&&{ \langle  \chi^{3 2}_{S_C M_{S_C}} \chi^{1 4}_{S_B M_{S_B}}|
\chi^{1 2}_{S_A M_{S_A}} \chi^{3 4}_{1 -\!m} \rangle} \nonumber\\&
=& \sum_{S_{BC},M_s}\langle  S_C , M_{S_C};S_B ,M_{S_B} |S,M_s
\rangle \langle S,M_s| S_A, M_{S_A};1, -m \rangle\;
\Big{[}3(2S_B+1)(2S_C+1)(2S_A+1)\Big{]}^{1/2} \left
\{\begin{array}{ccc}
1\over 2 & 1\over 2 & S_C \nonumber\\
1\over 2 & 1\over 2 & S_B\\
S_A & 1 & S
\end{array}
\right \} \;.
\end{eqnarray}
The relevant flavor matrix element is
\begin{eqnarray}
\langle \varphi^{3 2}_C \varphi^{1 4}_B| \varphi^{1 2}_A \varphi^{3
4}_0  \rangle = \sum_{I,I^{^{3}}}\langle
I_{C},I_{C}^{3};I_{B}I_{B}^{3}|I_{A},I_{A}^{3}\rangle[(2I_{B}+1)(2I_{C}+1)(2I_{A}+1)]^{1/2}
\left \{\begin{array}{ccc}
I_1 & I_3 & I_C \nonumber\\
I_2 & I_4 & I_B\\
I_A & 0 & I_{A}
\end{array}
\right \},
\end{eqnarray}
where $I_{i}\;(i=1,2,3,4)$ is the isospin of quark which is
labelled in Fig. \ref{fig1}.

With the Jacob-Wick formula the helicity amplitude can be converted
into the partial wave amplitude \cite{convert}
\begin{eqnarray}
{\mathcal{M}}^{J L}(A\rightarrow BC) = \frac{\sqrt{2 L+1}}{2 J_A +1}
\!\! \sum_{M_{J_B},M_{J_C}} \langle L, 0; J, M_{J_A}|J_A ,
M_{J_A}\rangle \langle J_B , M_{J_B}; J_C , M_{J_C} | J, M_{J_A}
\rangle \mathcal{M}^{M_{J_A} M_{J_B} M_{J_C}}({\textbf{k}}_B),
\end{eqnarray}
where $M_{J_A}=M_{J_B}+M_{J_C}$,
${\textbf{J}}={\textbf{J}}_B+{\textbf{J}}_C $ and
$\textbf{J}_{A}+\textbf{J}_{P}=\textbf{J}_{B}+\textbf{J}_{C}+\textbf{L}$.
The decay width in terms of partial wave amplitudes using the
relativistic phase space is
\begin{eqnarray*}
\Gamma = {1\over8\pi} \frac{{|\textbf{p}|}}{M_A^2}\sum_{JL}
|\mathcal{M}^{J L}|^2,
\end{eqnarray*}
where $|\textbf{p}|$ is the three momentum of the daughter mesons
in the parent's center of mass frame.

\section{Strong decays of $0^{+}(2^{3}P_{0})$ $c\bar{s}$
state}\label{sec3}

\subsection{$D^{0}K^{+},\; D^{+}K^{0},\; D_{s}^{+}\eta$ modes}

The harmonic oscillator wavefunctions of $0^{+}(2^{3}P_{0})$
$c\bar{s}$ state can be expressed as
\begin{eqnarray} \psi^{n=2;L=1}({{\mathbf{p}_1,\mathbf{p}_2}}) &=& \frac{iR^{5/2}}{\sqrt{15}\pi^{1/4}}
[10-{R^{2}(
{{\mathbf{p}_1-\mathbf{p}_2}})^{2}}]\mathcal{Y}_{1}^{m}\big(\frac{\mathbf{p}_{1}-\mathbf{p}_{2}}{2}\big)
\exp\Big{[}-{1\over 8}({{\mathbf{p}_1-\mathbf{p}_2})}^2R^2\Big{]}.
\end{eqnarray}
For $K$ and $\eta$ mesons, we use S-wave harmonic oscillator
wavefunction
\begin{eqnarray} \psi^{n=1;L=0}({{\mathbf{p}_1,\mathbf{p}_2}}) &=& \Big{(}\frac{R^2}{
\pi }\Big{)}^{3/4} \exp\Big{[}-{1\over
8}({{\mathbf{p}_1-\mathbf{p}_2})}^2R^2\Big{]},
\end{eqnarray}
where the parameter $R$ denotes the meson radius.

With the $^{3}P_{0}$ model, the general expression of the
amplitudes for the decay of $0^{+}(2^{3}P_{0})$ $c\bar{s}$ into
two pseudoscalar mesons reads
\begin{eqnarray}
\mathcal{M}(c\bar{s}(2^{3}P_{0})\to
0^{-}+0^{-})=\alpha\frac{\gamma\sqrt{2E_{A}E_{B}E_{C}}}{6\sqrt{3}}\big[I_{0}-2I_{\pm}\big]\label{factor-1}
\end{eqnarray}
with the spatial integral $I_{\pm}$ and $I_{0}$
\begin{eqnarray}
I_{\pm}&=&\frac{-2\sqrt{3}}{\sqrt{5}}\frac{{\pi^{1/4}}(R_{A}^{2})^{5/4}(R_{B}^{2})^{3/4}(R_{C}^{2})^{3/4}}{(R_{A}^{2}+R_{B}^{2}
+R_{C}^{2})^{9/2}}\big[10R_{A}^{4}+\mathbf{k}_{B}^{2}R_{A}^{2}(R_{B}^{2}+R_{C}^{2})^{2}-10(R_{B}^{2}+R_{C}^{2})^2\big]
\nonumber\\&&\times\exp\bigg[-\frac{\mathbf{k}_{B}^{2}R_{A}^{2}(R_{B}^{2}+R_{C}^{2})}{8(R_{A}^{2}+R_{B}^{2}+R_{C}^{2})}\bigg],\\
I_{0}&=&-2\sqrt{\frac{3}{5}}\frac{{\pi^{1/4}}(R_{A}^{2})^{5/4}(R_{B}^{2})^{3/4}(R_{C}^{2})^{3/4}}{(R_{A}^{2}+R_{B}^{2}+R_{C}^{2})
^{11/2}}\Big\{20R_{A}^{6}\big[(R_{B}^{2}+R_{C}^{2})\mathbf{k}_{B}^{2}-2\big]+2(R_{B}^{2}+R_{C}^{2})R_{A}^{4}
\big[\mathbf{k}_{B}^{4}(R_{B}^{2}+R_{C}^{2})^{2}\nonumber\\&&-(R_{B}^{2}+R_{C}^{2})\mathbf{k}_{B}^{2}-20\big]+
(R_{B}^{2}+R_{C}^{2})^{2}R_{A}^{2}\big[(R_{B}^{2}+R_{C}^{2})^{2}\mathbf{k}_{B}^{4}-32(R_{B}^{2}+R_{C}^{2})\mathbf{k}_{B}^{2}+40\big]\nonumber\\
&&-
10(R_{B}^{2}+R_{C}^{2})^{3}\big[\mathbf{k}_{B}^{2}(R_{B}^{2}+R_{C}^{2})-4\big]
\Big\}\exp\bigg[-\frac{\mathbf{k}_{B}^{2}R_{A}^{2}(R_{B}^{2}+R_{C}^{2})}{8(R_{A}^{2}+R_{B}^{2}+R_{C}^{2})}\bigg],
\end{eqnarray}
where indexes $A$, $B$ and $C$ correspond to $0^{+}(2^{3}P_{0})$
$c\bar{s}$ state, $D_{(s)}$ and $K(\eta)$ respectively. The factor
$\alpha$ in eq. (\ref{factor-1}) is $1$ for
$D^{0}K^{+}(D^{+}K^{0})$ modes and $2/\sqrt{6}$ for the
$D^{+}_{s}\eta$ mode.

\subsection{Double pion decays}

\begin{figure}[htb]
\begin{center}
\scalebox{0.8}{\includegraphics{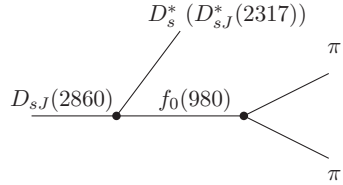}}
\end{center}
\caption{The Feynman diagram for the double pion decays of
$0^{+}(2^{3}P_{0})$ $c\bar{s}$ state.}\label{3872-two}
\end{figure}

$0^{+}(2^{3}P_{0})$ $c\bar{s}$ state can also decay into
$D^{*}_{s}\pi\pi$ and $D_{sJ}^{*}(2317)\pi\pi$. Such a decay could
occur via a virtual intermediate $f_{0}(980)$ or $\sigma$ meson,
which is shown in Fig. \ref{3872-two}. So we only consider the
contribution from $f_{0}(980)$ to make an estimate about the
double pion decay. The dominant decay mode of $f_{0}(980)$ is
$\pi\pi$. In this work, we assume $f_{0}(980)$ is composed of
$s\bar{s}$.

The effective Lagrangian describing $f_{0}(980)\to \pi\pi$ reads as
\begin{eqnarray}
\mathcal{L}_{f_{0}(980)\pi\pi}=g_{_{f_{0}\pi\pi}}[2\pi^{+}\pi^{-}+\pi^{0}\pi^{0}]f_{0},
\end{eqnarray}
where the coupling constant $g_{_{f_{0}\pi\pi}}$ taken as $0.83\sim
1.3$ GeV.

\subsubsection{$c\bar{s}(2^{3}P_{0})\to D_{s}^{*}f_{0}(980)\to
D_{s}^{*}\pi\pi$}

The amplitude of the $c\bar{s}(2^{3}P_{0})\to
D_{s}^{*}f_{0}(980)\to D_{s}^{*}\pi\pi$ decay chain can be
expressed as
\begin{eqnarray}
\mathcal{M}\big(c\bar{s}(2^{3}P_{0})\to D_{s}^{*}f_{0}(980)\to
D_{s}^{*}\pi\pi\big)=\mathcal{M}\big(c\bar{s}(2^{3}P_{0})\to
D_{s}^{*}f_{0}(980)\big)\frac{i}{k^{2}-m_{f_{0}}^{2}}\sqrt{\lambda_{\pi\pi}}g_{_{f_{0}\pi\pi}}
\end{eqnarray}
with $\lambda_{\pi^{+}\pi^{-}}=2$ and $\lambda_{\pi^{0}\pi^{0}}=1$.
$\mathcal{M}\big(c\bar{s}(2^{3}P_{0})\to D_{s}^{*}f_{0}(980)\big)$
can be calculated in the $^{3}P_{0}$ model using the wavefunction of
$f_{0}(980)$
\begin{eqnarray}
\psi^{n=1;L=1}({{\mathbf{p}_1,\mathbf{p}_2}})
&=&i\sqrt{\frac{2}{3}}\frac{R^{5/2}}{\pi^{1/4}}\mathcal{Y}_{1}^{m}(\mathbf{p}_{1}-\mathbf{p}_{2})
\exp\Big{[}-{1\over 8}({{\mathbf{p}_1-\mathbf{p}_2})}^2R^2\Big{]}.
\end{eqnarray}
Its expression reads
\begin{eqnarray}
\mathcal{M}\big(c\bar{s}(2^{3}P_{0})\to
D_{s}^{*}f_{0}(980)\big)=\frac{\gamma\sqrt{6E_{A}E_{B}E_{C}}}{9}\Big[
I^{0,0}_{0,0}+I^{1,-1}_{0,0}+I^{-1,1}_{0,0}
+I^{1,0}_{0,1}+I^{0,1}_{0,1}+I^{-1,0}_{0,-1}+I^{0,-1}_{0,-1}\Big],
\end{eqnarray}
where indices $A$, $B$ and $C$ correspond to $0^{+}(2^{3}P_{0})$
$c\bar{s}$ state, $D_{s}^{*}$ and $f_{0}(980)$. The expressions of
$I^{M_{LA},m}_{M_{LB},M_{LC}}$ are collected in Appendix A.

\subsubsection{ $c\bar{s}(2^{3}P_{0})\to
D_{sJ}^{*}(2317)f_{0}(980)\to D_{sJ}^{*}(2317)\pi\pi$}

We treat $D_{sJ}^{*}(2317)$ as a pure $c\bar{s}(0^{+})$ state.
Then we use $^{3}P_{0}$ model to make a very rough estimate of
$c\bar{s}(2^{3}P_{0})\to D_{sJ}^{*}(2317)f_{0}(980)\to
D_{sJ}^{*}(2317)\pi\pi$. Similarly, the amplitude of
$c\bar{s}(2^{3}P_{0})\to D_{sJ}^{*}(2317)f_{0}(980)\to
D_{sJ}^{*}(2317)\pi\pi$ decay chain reads as
\begin{eqnarray}
\mathcal{M}\big(c\bar{s}(2^{3}P_{0})\to
D_{sJ}^{*}(2317)f_{0}(980)\to
D_{sJ}^{*}(2317)\pi\pi\big)=\mathcal{M}\big(c\bar{s}(2^{3}P_{0})\to
D_{sJ}^{*}(2317)f_{0}(980)\big)\frac{i}{k^{2}-m_{f_{0}}^{2}}\sqrt{\lambda_{\pi\pi}}g_{_{f_{0}\pi\pi}},
\end{eqnarray}
where
\begin{eqnarray}
\mathcal{M}\big(c\bar{s}(2^{3}P_{0})\to
D_{sJ}^{*}(2317)f_{0}(980)\big)&=&\frac{\gamma\sqrt{2E_{A}E_{B}E_{C}}}{9\sqrt{3}}
\Big[I^{0,0}_{0,0}+I^{0,1}_{0,1}+I^{0,1}_{1,0}+I^{1,0}_{0,1}+I^{1,0}_{1,0}+I^{0,-1}_{0,-1}+I^{0,-1}_{-1,0}
+I^{-1,0}_{0,-1}\nonumber\\&&+I^{-1,0}_{-1,0}
+I^{1,-1}_{0,0}+I^{-1,1}_{0,0}+I^{0,0}_{1,-1}+I^{0,0}_{-1,1}+2I^{1,1}_{1,1}+2I^{-1,-1}_{-1,-1}\Big].
\end{eqnarray}
Indices $A$, $B$ and $C$ correspond to $0^{+}(2^{3}P_{0})$
$c\bar{s}$ state, $D_{sJ}^{*}(2317)$ and $f_{0}(980)$
respectively. The lengthy expressions of
$I^{M_{L_A},m}_{M_{L_B},M_{L_C}}$ are collected in Appendix A.

The three-body decay width of $A\to BCD$ reads \cite{PDG}
\begin{eqnarray}
\Gamma=\frac{1}{(2\pi)^{3}}\frac{1}{32M_{A}^{2}}\int|\mathcal{M}|^{2}dm_{12}^{2}dm_{23}^{2}.
\end{eqnarray}

\section{Strong decays of $1^{-}(1^{3}D_{1})$ $c\bar{s}$ state}
\label{sec4}

\subsection{$D^{0}K^{+},\;D^{+}K^{0},\;D_{s}^{+}\eta$ modes}

Using the harmonic-oscillator wavefunction of $1^{-}(1^{3}D_{1})$
$c\bar{s}$ state
\begin{eqnarray} \psi^{n=1;L=2}({{\mathbf{p}_1,\mathbf{p}_2}}) &=& \frac{R^{7/2}}{\sqrt{15}\pi^{1/4}}
\mathcal{Y}_{2}^{m}\big(\frac{\mathbf{p}_{1}-\mathbf{p}_{2}}{2}\big)
\exp\Big{[}-{1\over 8}({{\mathbf{p}_1-\mathbf{p}_2})}^2R^2\Big{]}
\end{eqnarray}
we have the decay amplitude
\begin{eqnarray}
\mathcal{M}\big(c\bar{s}(1^{3}D_{1})\to
0^{-}+0^{-}\big)=\alpha\frac{\gamma\sqrt{8E_{A}E_{B}E_{C}}}{3}\big[-\frac{1}{\sqrt{30}}I^{0,0}
+\frac{1}{2\sqrt{10}}I^{1,-1}+\frac{1}{2\sqrt{10}}I^{-1,1}\big]
\end{eqnarray}
where
\begin{eqnarray}
I^{0,0}&=&\frac{|\mathbf{k}_{B}|\pi^{1/4}R_{A}^{7/2}R_{B}^{3/2}R_{C}^{3/2}(\xi-1)}{(R_{A}^{2}+R_{B}^{2}+R_{C}^{2})^{5/2}}\bigg[
(R_{A}^{2}+R_{B}^{2}+R_{C}^{2})(\xi^{2}-1)\mathbf{k}_{B}^{2}+8\bigg]
\exp\bigg[-\frac{\mathbf{k}_{B}^{2}R_{A}^{2}(R_{B}^{2}+R_{C}^{2})}{8(R_{A}^{2}+R_{B}^{2}+R_{C}^{2})}\bigg],\label{I00}\\
I^{1,-1}&=&I^{-1,1}=-\frac{4\sqrt{3}|\mathbf{k}_{B}|\pi^{1/4}R_{A}^{7/2}R_{B}^{3/2}R_{C}^{3/2}(\xi-1)}{(R_{A}^{2}
+R_{B}^{2}+R_{C}^{2})^{5/2}}\exp\bigg[-\frac{\mathbf{k}_{B}^{2}R_{A}^{2}(R_{B}^{2}+R_{C}^{2})}
{8(R_{A}^{2}+R_{B}^{2}+R_{C}^{2})}\bigg]\label{I11}.
\end{eqnarray}
The indices $A$, $B$ and $C$ correspond to $1^{-}(1^{3}D_{1})$
$c\bar{s}$ state, $D_{(s)}$ and $K(\eta)$ respectively.

\subsection{$D^{*}K,\;DK^{*},\;D_{s}^{*+}\eta$ modes}

The $1^{-}(1^{3}D_{1})$ $c\bar{s}$ state can also decay into
$D^{*}K,\;D K^{*},\;D_{s}^{*+}\eta$ modes. The general decay
amplitude can be expressed as
\begin{eqnarray}
\mathcal{M}\big(c\bar{s}(1^{3}D_{1})\to
1^{-}+0^{-}\big)=\alpha\frac{\gamma\sqrt{8E_{A}E_{B}E_{C}}}{3\sqrt{2}}\big[\frac{1}{\sqrt{30}}I^{0,0}
+\frac{1}{2\sqrt{10}}I^{1,1}+\frac{1}{2\sqrt{10}}I^{-1,-1}\big],
\end{eqnarray}
where $I^{0,0}$, $I^{1,-1}$ and $I^{-1,1}$ are same as those in
Eqs. (\ref{I00}) and (\ref{I11}).

\subsection{Double pion decays}

\subsubsection{$D_{sJ}(2860)\to D_{sJ}^{*}(2317)f_{0}(980)\to
D_{sJ}^{*}(2317)\pi\pi$}

The amplitude of the $c\bar{s}(1^{3}D_{1})\to
D_{sJ}^{*}(2317)f_{0}(980)\to D_{sJ}^{*}(2317)\pi\pi$ decay chain is
written as
\begin{eqnarray}
\mathcal{M}\big(c\bar{s}(1^{3}D_{1})\to
D_{sJ}^{*}(2317)f_{0}(980)\to
D_{sJ}^{*}(2317)\pi\pi\big)=\mathcal{M}\big(c\bar{s}(1^{3}D_{1})\to
D_{sJ}^{*}(2317)f_{0}(980)\big)\frac{i}{k^{2}-m_{f_{0}}^{2}}\sqrt{\lambda_{\pi\pi}}g_{_{f_{0}\pi\pi}},
\end{eqnarray}
where
\begin{eqnarray}
\mathcal{M}(c\bar{s}(1^{3}D_{1})\to
D_{sJ}^{*}(2317)f_{0}(980))&=&\frac{\gamma\sqrt{8E_{A}E_{B}E_{C}}}{3}\bigg[
\frac{1}{6\sqrt{10}}\Big(I^{-1,0}_{-1,0}+I^{-1,0}_{0,-1}+I^{-1,1}_{0,0}+I^{1,-1}_{0,0}+I^{1,0}_{0,1}+I^{1,0}_{1,0}\Big)
\nonumber\\&&+\frac{1}{3\sqrt{30}}\Big(I^{0,-1}_{-1,0}+I^{0,-1}_{0,-1}+I^{0,0}_{-1,1}+I^{0,0}_{0,0}+I^{0,0}_{1,-1}+I^{0,1}_{0,1}+I^{0,1}_{1,0}\Big)
\nonumber\\&&+\frac{1}{3\sqrt{10}}\Big(I^{-1,-1}_{-1,-1}+I^{1,1}_{1,1}\Big)
\bigg].\label{express-1}
\end{eqnarray}
The indices $A$, $B$ and $C$ correspond to $c\bar{s}(1^{3}D_{1})$
state, $D_{sJ}^{*}(2317)$ and $f_{0}(980)$.

\subsubsection{$c\bar{s}(1^{3}D_{1})\to D_{s}^{*}f_{0}(980)\to
D_{s}^{*}\pi\pi$}

The amplitude of the $c\bar{s}(1^{3}D_{1})\to
D_{s}^{*}f_{0}(980)\to D_{s}^{*}\pi\pi$ decay chain is
\begin{eqnarray}
\mathcal{M}\big(c\bar{s}(1^{3}D_{1})\to D_{s}^{*}f_{0}(980)\to
D_{s}^{*}\pi\pi\big)=\mathcal{M}\big(c\bar{s}(1^{3}D_{1})\to
D_{s}^{*}f_{0}(980)\big)\frac{i}{k^{2}-m_{f_{0}}^{2}}\sqrt{\lambda_{\pi\pi}}g_{_{f_{0}\pi\pi}},
\end{eqnarray}
with
\begin{eqnarray}
\mathcal{M}(c\bar{s}(1^{3}D_{1})\to
D_{s}^{*}f_{0}(980))&=&\frac{\sqrt{5}\gamma\sqrt{8E_{A}E_{B}E_{C}}}{3\sqrt{3}}\bigg[
\frac{\sqrt{3}}{20}\Big(I^{-1,0}_{0,-1}+I^{-1,1}_{0,0}+I^{1,-1}_{0,0}+I^{1,0}_{0,1}\Big)
+\frac{1}{10} \Big(  I^{0,-1}_{0,-1} +I^{0,0}_{0,0}+I^{0,1}_{0,1}
\Big) \bigg],\nonumber\\\label{express-2}
\end{eqnarray}
where indexes $A$, $B$ and $C$ correspond to
$c\bar{s}(1^{3}D_{1})$ state, $D_{s}^{*}$ and $f_{0}(980)$
respectively. We collect the detailed expressions of
$I^{M_{L_A},m}_{M_{L_B},M_{L_C}}$ in Eqs. (\ref{express-1}) and
(\ref{express-2}) in Appendix B.

\section{Strong decays of $1^{-}(2^{3}S_{1})$ $c\bar{s}$ state}
\label{sec5}

\subsection{$D^{0}K^{+},\;D^{+}K^{0},\;D_{s}^{+}\eta$ modes}

Using the harmonic oscillator wavefunction of the
$1^{-}(2^{3}S_{1})$ $c\bar{s}$ state
\begin{eqnarray}
\psi^{n=2;L=0}({{\mathbf{p}_1,\mathbf{p}_2}}) &=&
\frac{1}{\sqrt{4\pi}}\bigg (\frac{4R^3}{\sqrt{\pi}}\bigg)^{{1}/{2}}
\sqrt{\frac{2}{3}} \bigg[\frac{3}{2} - \frac{R^2}{4}
 ({{\mathbf{p}_1-\mathbf{p}_2})}^2 \bigg]\exp\Big{[}-{1\over 8}({{\mathbf{p}_1-\mathbf{p}_2})}^2
 R^2\Big{]},
 \nonumber
\end{eqnarray}
the general decay amplitude of $c\bar{s}(2^{3}S_{1})\to
0^{-}+0^{-}$ can be written as
\begin{eqnarray}
\mathcal{M}(c\bar{s}(2^{3}S_{1})\to
0^{-}+0^{-})&=&\frac{\gamma\sqrt{8E_{A}E_{B}E_{C}}}{\sqrt{3}} \bigg(
\frac{1}{6} I^{0,0}_{0,0}  \bigg) \nonumber
\end{eqnarray}
where
\begin{eqnarray} I^{0,0}_{0,0}&=& - \sqrt{\frac{1}{2}}
\frac{|\mathbf{k}_{B}|
    \pi^{1/4}R_{A}^{3/2}R_{B}^{3/2}R_{C}^{3/2}}
{(R_{A}^{2}+R_{B}^{2}+R_{C}^{2})^{5/2}}             \Bigg\{
-6(R_{A}^{2}+R_{B}^{2}+R_{C}^{2})(1+\xi)+R_A^2  \bigg[
4+20\xi\nonumber\\&&+\mathbf{k}_{B}^2
(R_{A}^{2}+R_{B}^{2}+R_{C}^{2}) (-1+\xi)^2(1+\xi) \big]
 \Bigg\}
\exp\bigg[-\frac{\mathbf{k}_{B}^{2}R_{A}^{2}(R_{B}^{2}+R_{C}^{2})}{8(R_{A}^{2}+R_{B}^{2}+R_{C}^{2})}\bigg],
\label{express-3}
\end{eqnarray}
The indices $A$, $B$ and $C$ denote the $c\bar{s}(2^{3}S_{1})$
state, $D_{(s)}$ and $K(\eta)$ respectively.

\subsection{$D^{*}K,\;DK^{*},\;D_{s}^{*+}\eta$ modes}

Similarly, one can  get
\begin{eqnarray}
\mathcal{M}(c\bar{s}(2^{3}S_{1})\to 1^{-}+0^{-}
)&=&\frac{\gamma\sqrt{8E_{A}E_{B}E_{C}}}{\sqrt{3}} \bigg(
\frac{1}{3 \sqrt{2}} I^{0,0}_{0,0}  \bigg),
\end{eqnarray}
where the expression of $I^{0,0}_{0,0}$ is same as that in Eq.
(\ref{express-3}).

\subsection{Double pion decays}

\subsubsection{$c\bar{s}(2^{3}S_{1})\to
D_{sJ}^{*}(2317)f_{0}(980)\to D_{sJ}^{*}(2317)\pi\pi$}

The amplitude of the $c\bar{s}(2^{3}S_{1})\to
D_{sJ}^{*}(2317)f_{0}(980)\to D_{sJ}^{*}(2317)\pi\pi$ decay chain
is
\begin{eqnarray}
\mathcal{M}\big(c\bar{s}(2^{3}S_{1})\to
D_{sJ}^{*}(2317)f_{0}(980)\to
D_{sJ}^{*}(2317)\pi\pi\big)=\mathcal{M}\big(c\bar{s}(2^{3}S_{1})\to
D_{sJ}^{*}(2317)f_{0}(980)\big)\frac{i}{k^{2}-m_{f_{0}}^{2}}\sqrt{\lambda_{\pi\pi}}g_{_{f_{0}\pi\pi}}
\end{eqnarray}
with
\begin{eqnarray}
\mathcal{M}(c\bar{s}(2^{3}S_{1})\to
D_{sJ}^{*}(2317)f_{0}(980))&=&\frac{\gamma\sqrt{8E_{A}E_{B}E_{C}}}{\sqrt{3}}\bigg[
-\frac{1}{18} \Big( I^{0,-1}_{-1,0} + I^{0,-1}_{0,-1} +
I^{0,0}_{-1,1} + I^{0,0}_{0,0} + I^{0,0}_{1,-1} + I^{0,1}_{0,1} +
I^{0,1}_{1,0}    \Big) \bigg],\nonumber\\\label{expression-4}
\end{eqnarray}
where the indices $A$, $B$ and $C$ correspond to the
$c\bar{s}(2^{3}S_{1})$ state, $D_{sJ}^{*}(2317)$ and $f_{0}(980)$.

\subsubsection{$c\bar{s}(2^{3}S_{1})\to D_{s}^{*}f_{0}(980))\to
D_{s}^{*}\pi\pi$}

The amplitude of the $c\bar{s}(2^{3}S_{1})\to
D_{s}^{*}f_{0}(980)\to D_{s}^{*}\pi\pi$ decay chain is
\begin{eqnarray}
\mathcal{M}\big(c\bar{s}(2^{3}S_{1})\to D_{s}^{*}f_{0}(980)\to
D_{s}^{*}\pi\pi\big)=\mathcal{M}\big(c\bar{s}(2^{3}S_{1})\to
D_{s}^{*}f_{0}(980)\big)\frac{i}{k^{2}-m_{f_{0}}^{2}}\sqrt{\lambda_{\pi\pi}}g_{_{f_{0}\pi\pi}}
\end{eqnarray}
with
\begin{eqnarray}
\mathcal{M}(c\bar{s}(2^{3}S_{1})\to D_{s}^{*}
f_{0}(980))&=&\frac{\gamma\sqrt{8E_{A}E_{B}E_{C}}}{3}\bigg[
\frac{1}{2 \sqrt{3}}\Big( I^{0,-1}_{0,-1} + I^{0,0}_{0,0}
+I^{0,1}_{0,1}  \Big) \bigg],\label{expression-5}
\end{eqnarray}
where the indices $A$, $B$ and $C$ correspond to
$c\bar{s}(2^{3}S_{1})$ state, $D_{s}^{*}$ and $f_{0}(980)$. We
collect the expressions of $I^{M_{L_A},m}_{M_{L_B},M_{L_C}}$ in
Eqs. (\ref{expression-4}) and (\ref{expression-5}) in Appendix C

\section{Strong decays of $3^{-}(1^{3}{D}_{3})$ $c\bar{s}$
state}\label{sec6}

\subsection{$D^{0}K^{+}, D^{+}K^{0}, D_{s}\eta$ modes}

The general expression of the decay amplitude for reads
$c\bar{s}(1^{3}D_{3})\to 0^{-}+0^{-}$
\begin{eqnarray}
\mathcal{M}\big(c\bar{s}(1^{3}D_{3})\to
0^{-}+0^{-}\big)=\alpha\frac{\gamma\sqrt{8E_{A}E_{B}E_{C}}}{\sqrt{21}}\bigg[\frac{1}{2\sqrt{5}}I^{0,0}+
\frac{1}{2\sqrt{15}}I^{1,-1}+ \frac{1}{2\sqrt{15}}I^{-1,1}\bigg],
\end{eqnarray}
where the expressions of $I^{0,0}$, $I^{1,-1}$ and $I^{-1,1}$ are
also same as those in Eqs. (\ref{I00}) and (\ref{I11}). The
indices of $A$, $B$ and $C$ correspond to $3^{-}(1^{3}D_{3})$
$c\bar{s}$ state, $D_{(s)}$ and $K(\eta)$ respectively.

\subsection{$D^{*}K,\;DK^{*},\;D_{s}^{*+}\eta$ modes}

The general decay amplitude for $c\bar{s}(1^{3}D_{3})\to
1^{-}+0^{-}$ can be expressed as
\begin{eqnarray}
\mathcal{M}\big(c\bar{s}(1^{3}D_{3})\to
1^{-}+0^{-}\big)=\alpha\frac{\gamma\sqrt{8E_{A}E_{B}E_{C}}}{\sqrt{42}}\big[\frac{2}{\sqrt{30}}I^{0,0}
+\frac{1}{3}\sqrt{\frac{2}{5}}I^{1,-1}+\frac{1}{3}\sqrt{\frac{2}{5}}I^{-1,1}\big],
\end{eqnarray}
where the expressions of $I^{0,0}$, $I^{1,-1}$ and $I^{-1,1}$ are
same as those in Eqs. (\ref{I00}) and (\ref{I11}).

\section{Numerical results}\label{sec7}

The calculation of transition amplitude using $^{3}P_{0}$ model
involves two parameters: the strength of quark pair creation from
vacuum $\gamma$ and the $R$ value in the harmonic oscillator
wavefunction. We follow the convention of Ref. \cite{Godfrey} and
take $\gamma=6.9$, which is $\sqrt{96\pi}$ times larger than that
used by other groups \cite{close-3p0,kokoski}. $\gamma$ is a
universal parameter in the $^{3}P_{0}$ model.

Throughout our calculations of the decay amplitudes, we use the
harmonic oscillator wavefunctions as commonly done in the
framework of the $^{3}P_{0}$ model in literature, which indeed
brings some uncertainties. A more realistic wavefunction like that
from the linear potential model seems more reliable to describe
the properties of the meson. But the decay amplitude with more
realistic wavefunctions does not have the simple analytical
expression. Moreover such a more complicated calculation with more
realistic wavefunctions does not always lead to systematic
improvements because of the inherent uncertainties of the
$^{3}P_{0}$ srong decay model itself as discussed in Ref.
\cite{qpc-1}. The value of the parameter R in the harmonic
oscillator wavefunction can be fixed to reproduce the realistic
root mean square (RMS) radius, which is obtained by solving the
schr\"{o}dinger equation with the linear potential.

The R values in the harmonic oscillator wavefunctions are taken from
Ref. \cite{close-3p0}. We collect all these values in Table
\ref{table-1}. The masses of $K^{(*)}$, $\eta$, $f_{0}(980)$ and
charm-strange mesons are taken from PDG \cite{PDG}.

\begin{table}[htb]
\begin{center}
\begin{tabular}{c|ccccccccc} \hline
&$K$&$K^{*}$&$\eta$&$f_{0}(980)$&$D(0^{-})$&$D_{s}(0^{-})$&$D^{*}(1^{-})$&$D_{s}^{*}(1^{-})$&$D_{sJ}^{*}(2317)$\\\hline\hline
mass
(GeV)&0.494&0.892&0.548&0.980&1.869&1.968&2.007&2.112&2.317\\\hline
R (GeV$^{-1}$)&2.17&3.13&2.08&2.78&2.33&1.92&2.70&2.22&2.70
\\\hline
\end{tabular}
\end{center}
\caption{The meson masses and R values used in our calculation.
\label{table-1}}
\end{table}

\begin{figure}[htb]
\begin{center}
\scalebox{0.9}{\includegraphics{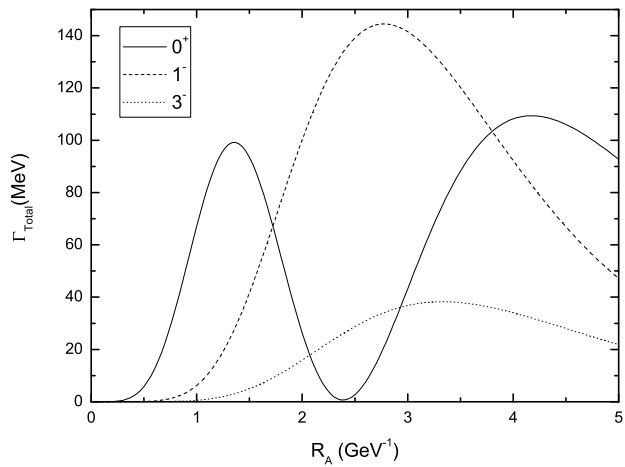}}
\end{center}
\caption{The dependence of $D_{sJ}(2860)$'s total two-body decay
widths on $R_{A}$ and possible quantum numbers.}\label{two-body}
\end{figure}

The dependence of $D_{sJ}(2860)$'s total two-body decay widths on
its $R_{A}$ and its possible quantum numbers is presented in Fig.
\ref{two-body}. As expected, its width vanishes around $R_A=2.4$
GeV$^{-1}$ for the $0^{+}(2^{3}P_{0})$ case because of the node in
the radial wavefunction.

\begin{figure}[htb]
\begin{center}
\begin{tabular}{cc}
\scalebox{0.8}{\includegraphics{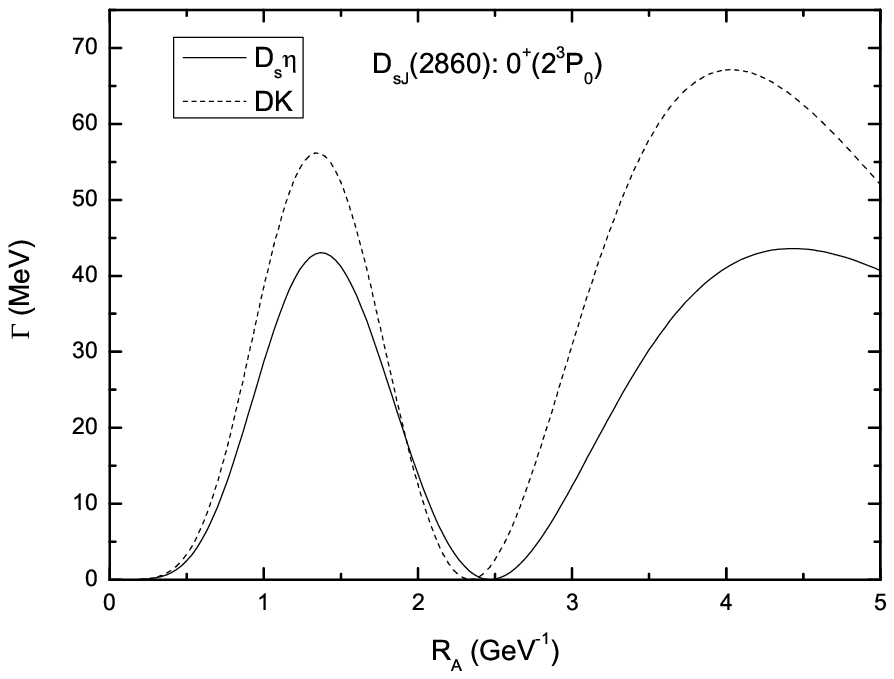}}&\scalebox{0.8}{\includegraphics{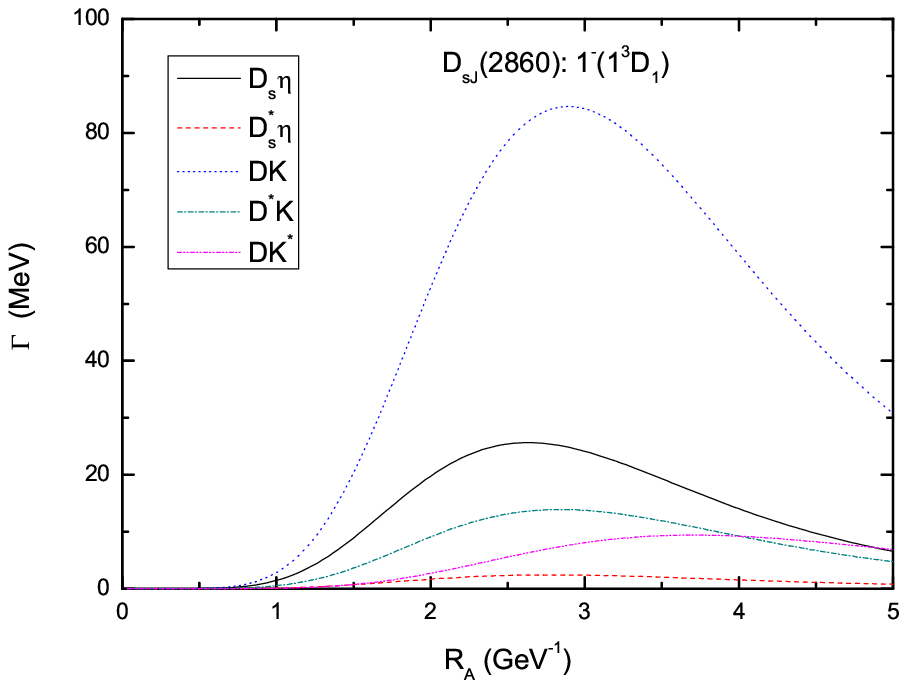}}\\
(a)&(b)\\
\scalebox{0.8}{\includegraphics{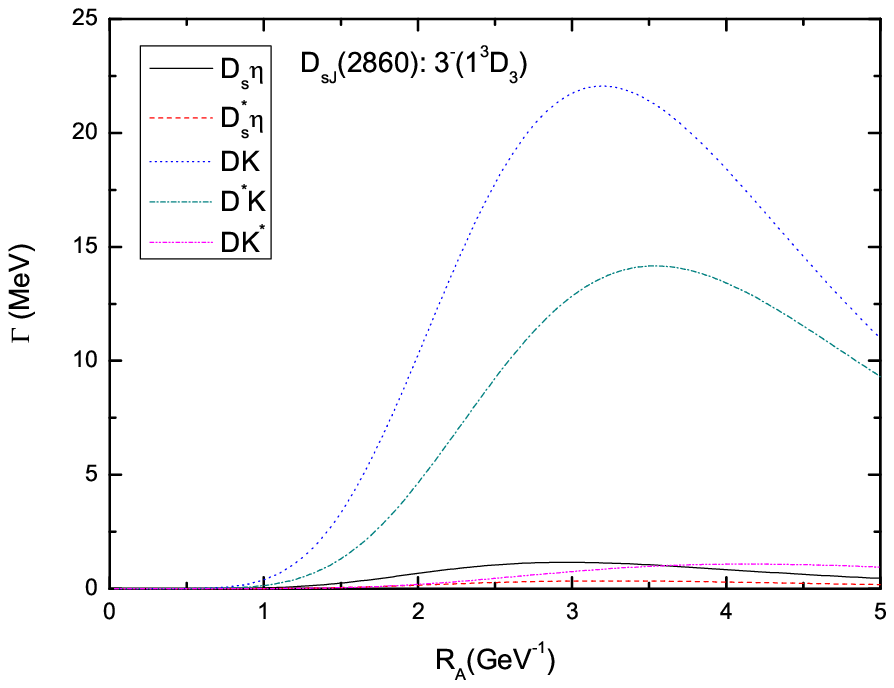}}&\scalebox{0.8}{\includegraphics{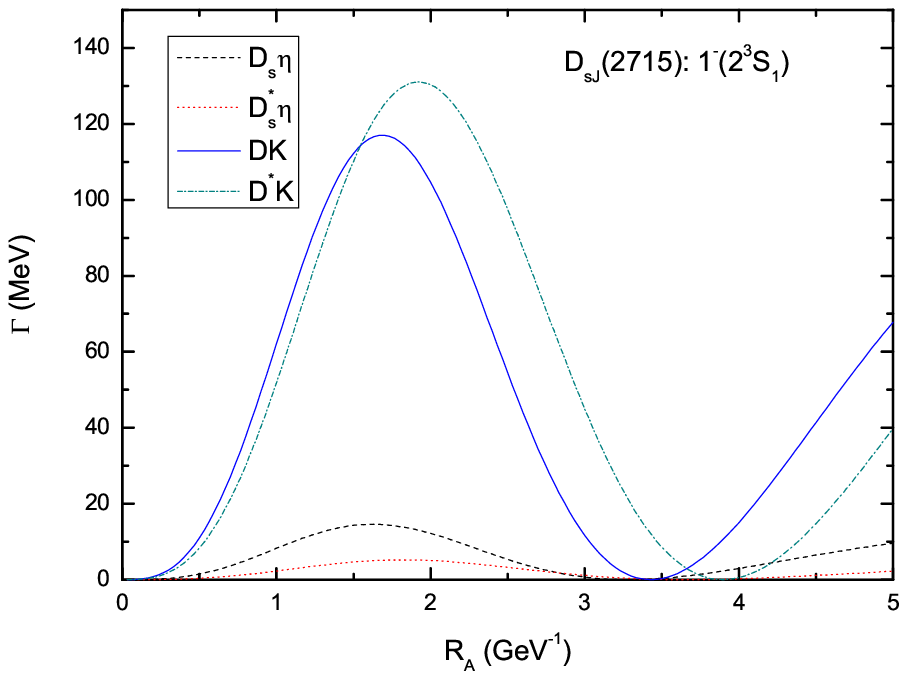}}\\
(c)&(d)
\end{tabular}
\end{center}
\caption{(a) The dependence of $D_{sJ}(2860)$'s two-body decay
widths on $R_{A}$ as a $0^{+}(2^{3}P_{0})$ candidate; (b) For the
case of $1^{-}(1^{3}D_{1})$; (c) For the case of
$3^{-}(1^{3}D_{3})$; (d) For the case of $D_{sJ}(2715)$ as a
$1^{-}(2^{3}S_{1})$ candidate. \label{2-body-0}}
\end{figure}

The variation of the decay widths of $D_{sJ}(2860)$'s and
$D_{sJ}(2715)$'s different modes with $R_A$ and their quantum
numbers is shown in Fig. \ref{2-body-0}.

\begin{figure}[htb]
\begin{center}
\begin{tabular}{cc}
\scalebox{0.8}{\includegraphics{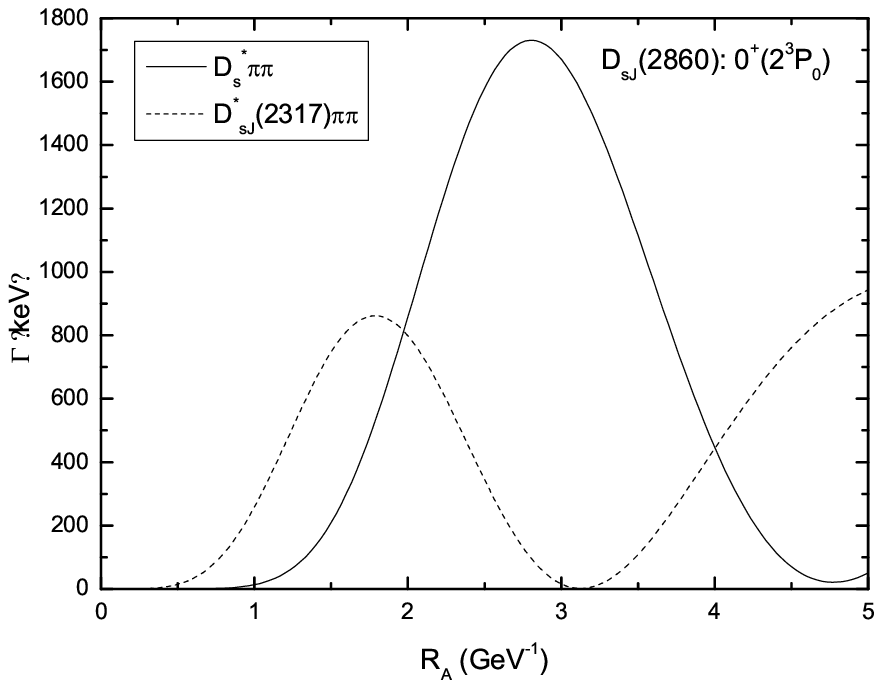}}&\scalebox{0.8}{\includegraphics{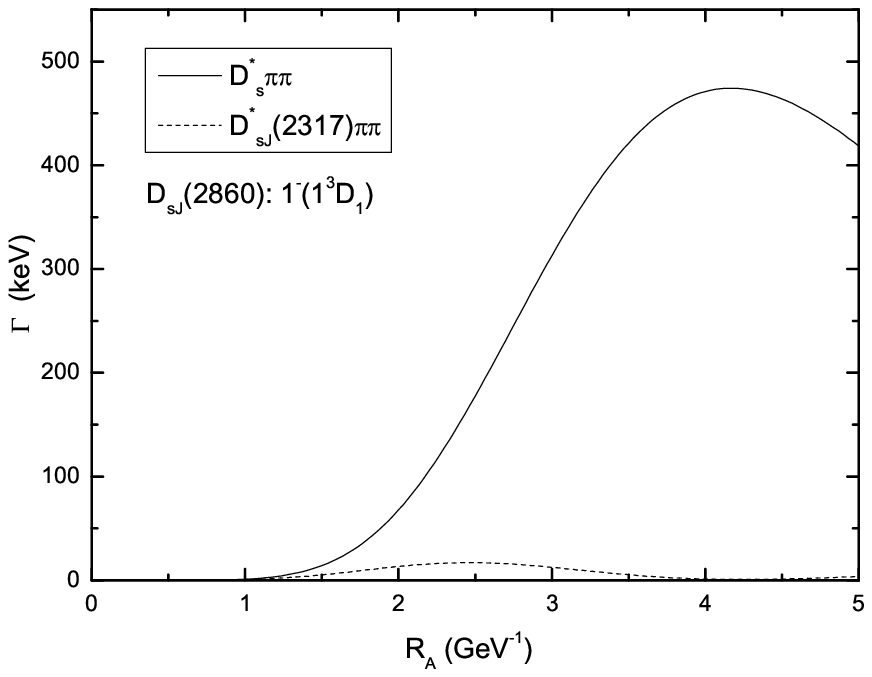}}\\
(a)&(b)
\end{tabular}
\end{center}
\caption{(a) The dependence of $D_{sJ}(2860)$'s three-body decay
widths on $R_{A}$ as a $0^{+}$ $c\bar{s}$ state. (b) As a
$1^{-}(^3D_1)$ $c\bar{s}$ state.}\label{three-body-0}
\end{figure}

The variation of $D_{sJ}(2860)$'s three-body decay widths with
$R_A$ is presented in Fig. \ref{three-body-0}.

\begin{figure}[htb]
\begin{center}
\begin{tabular}{cc}
\scalebox{0.8}{\includegraphics{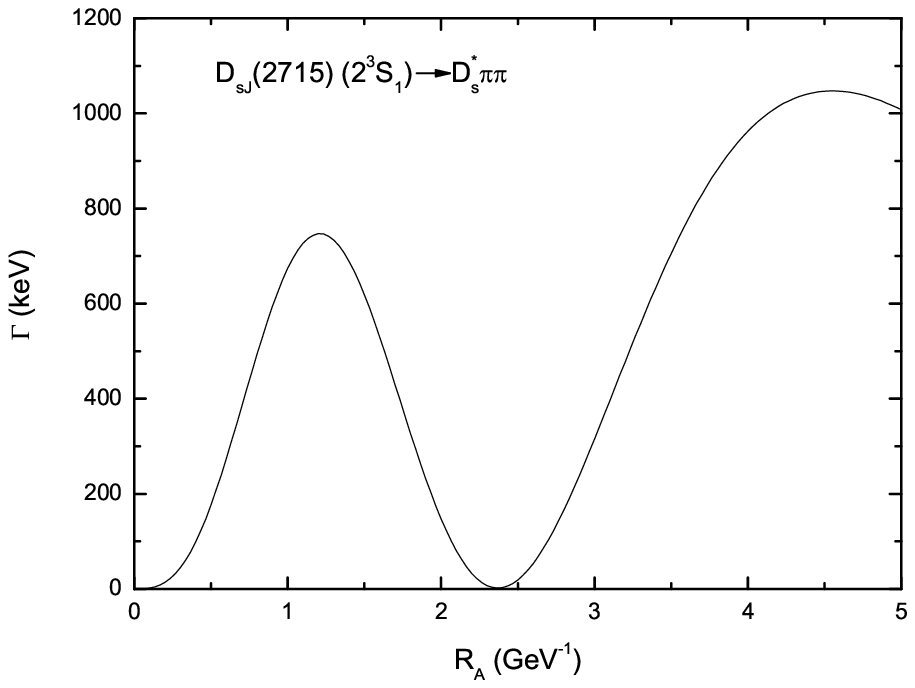}}&\scalebox{0.8}{\includegraphics{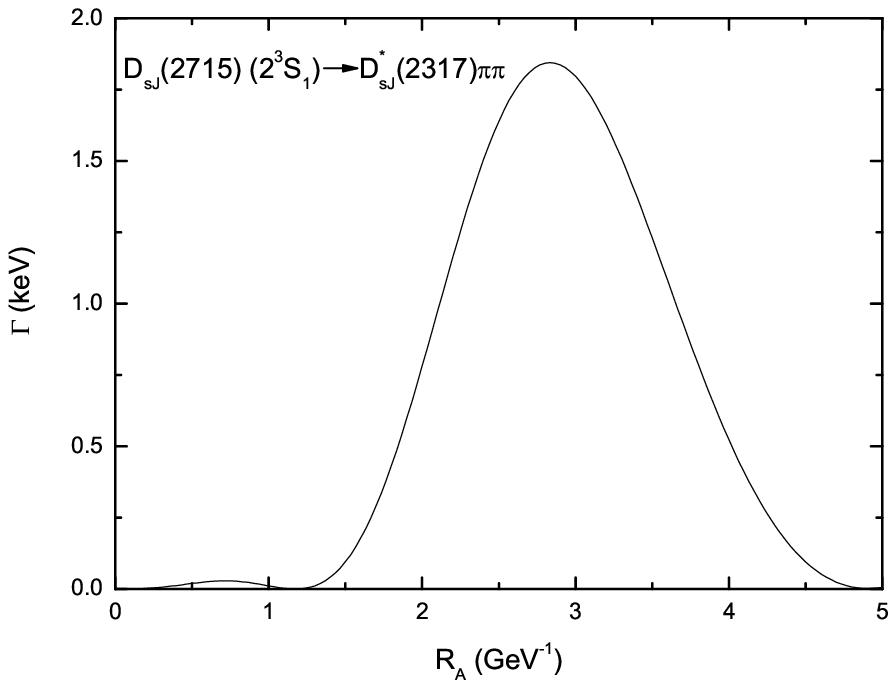}}\\
(a)&(b)
\end{tabular}
\end{center}
\caption{$D_{sJ}(2715)$'s three-body decay widths as $2^{3}S_{1}$
$c\bar{s}$.\label{2715-3-body}}
\end{figure}

The variation of $D_{sJ}(2715)$'s different three-body decay
widths with $R_A$ is presented in Fig. \ref{2715-3-body}

\begin{figure}[htb]
\begin{center}
\begin{tabular}{cc}
\scalebox{0.8}{\includegraphics{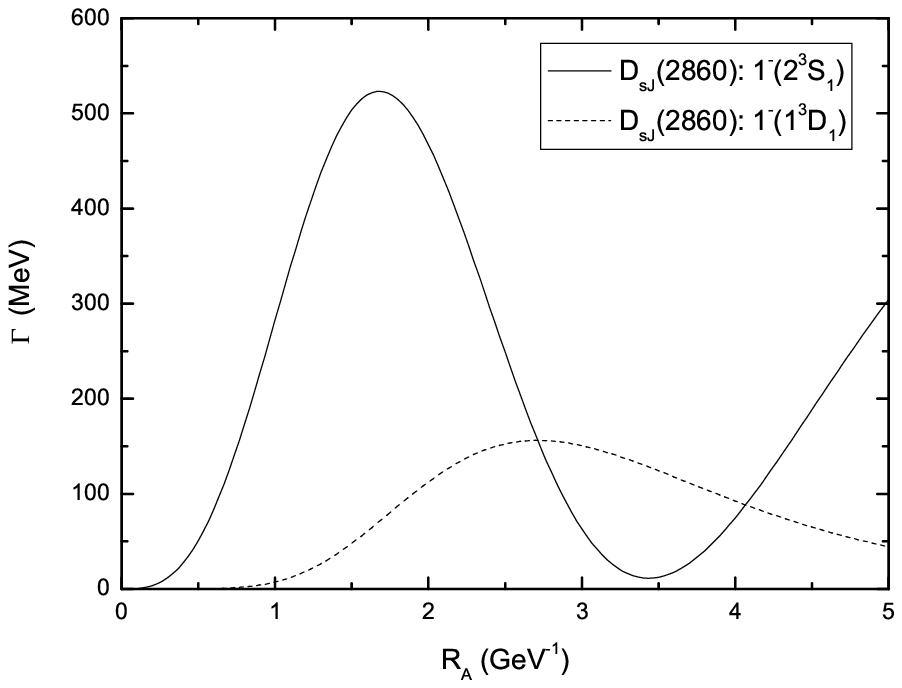}}&\scalebox{0.8}{\includegraphics{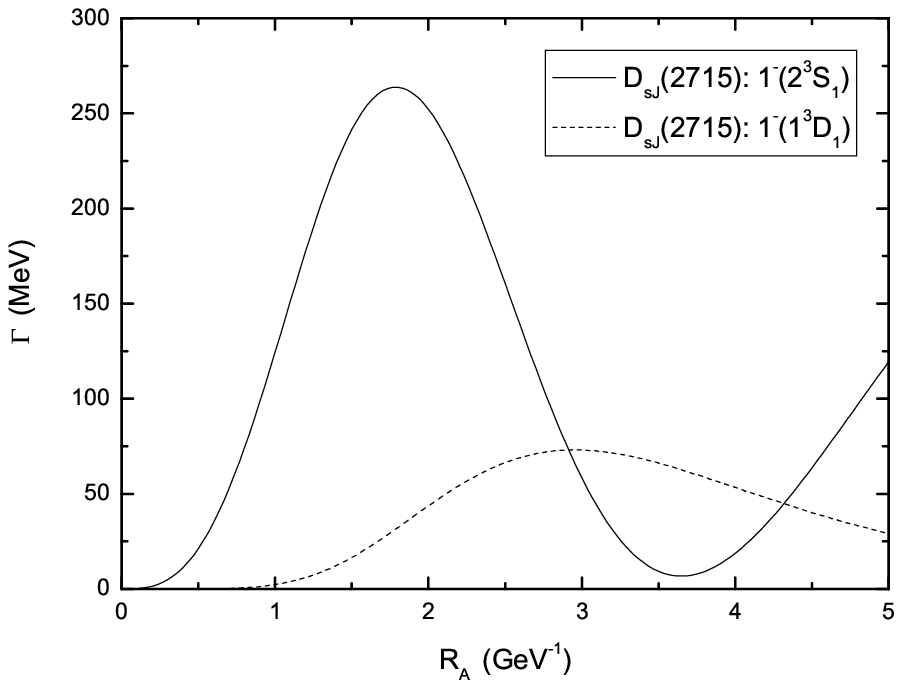}}\\
(a)&(b)
\end{tabular}
\end{center}
\caption{The variation of both $D_{sJ}(2860)$ and $D_{sJ}(2715)$'s
two-body decay widths with $R_A$ as either a $1^{3}D_{1}$ state or
$2^{3}S_{1}$ state.\label{2860-2715}}
\end{figure}

We present the variation of both $D_{sJ}(2860)$ and
$D_{sJ}(2715)$'s two-body decay widths with $R_A$ as either a
$1^{3}D_{1}$ state or $2^{3}S_{1}$ state in Fig. \ref{2860-2715}.

We take $R_{A}=3.1$ GeV$^{-1}$ for $0^{+}(2^{3}P_{1})$,
$R_{A}=3.2$ GeV$^{-1}$ for $1^{-}(2^{3}S_{1})$ $c\bar s$ states
and $R_{A}=2.94$ GeV$^{-1}$ for $1^{-}(1^{3}D_{1})$ and
$3^{-}(1^{3}D_{3})$ $c\bar s$ states to estimate the decay widths
and collect them in Table \ref{table-2}, which is the main result
of this work.
\begin{table}[htb]
\begin{center}
\begin{tabular}{c|ccccccccccc} \hline
&&&&&$D_{sJ}(2860)$\\\hline &$DK$ &$D_{s}\eta$ &$D^{*}K$
&$D_{s}^{*}\eta$  &$DK^*$   &$D_{s}^{*}\pi\pi$
&$D_{sJ}^{*}(2317)\pi\pi$& Total
\\\hline $0^{+}(2^{3}P_{0})$&37&16&-&-&-&1600&2&54\\\hline
$1^{-}(1^{3}D_{1})$&84&24&14&2&7.8&313&13&132\\\hline
$1^{-}(2^{3}S_{1})$&0.012&0.104&24&1.6&64&923&30&90\\\hline
$3^{-}(1^{3}D_{3})$& 22&1.2&13&0.3&0.71&-&-&37\\\hline\hline
&&&&&$D_{sJ}(2715)$\\\hline&$DK$ &$D_{s}\eta$&$D^{*}K$
&$D_{s}^{*}\eta$& &$D_{s}^{*}\pi\pi$ &$D_{sJ}^{*}(2317)\pi\pi$
&Total\\\hline $1^- (2^3 S_1)$ &3.2&0.05&27.2&0.54&&477&1.63&32
\\\hline
$1^- (1^3 D_1)$ &49.4&13.2&8&2.4&&49&0.5&73\\\hline
\end{tabular}
\end{center}
\caption{The decay widths of $D_{sJ}(2860)$ and $D_{sJ}(2715)$ as
different $c\bar s$ candidates. The two-body decay width and total
width are in unit of MeV while the three-body decay width is in
unit of keV. \label{table-2}}
\end{table}

\section{Discussions}\label{sec8}

We have studied the strong decay patterns of $D_{sJ}(2860)$ and
$D_{sJ}(2715)$ using $^{3}P_{0}$ model, assuming they are excited
$c\bar s$ candidates . If $D_{sJ}(2715)$ is the
$1^{-}(2^{3}S_{1})$ candidate, $D^{*}K$ should be the dominant
decay mode with a width around 27 MeV. In contrast, the decay
width of the DK mode is only 3 MeV. The $D_s^\ast \pi\pi$ mode has
a rather large width around 0.48 GeV. Its total width is only 32
MeV, much smaller than the experimental value $\Gamma=(115\pm
20^{+36}_{-32})$ MeV. However, one should be cautious that the
total width is sensitive to $R_A$ and the node position. On the
other hand, DK becomes the dominant decay mode and its total width
is around 73 MeV, roughly consistent with experimental data if one
identifies $D_{sJ}(2715)$ as the $1^{-}(1^{3}D_{1})$ $c\bar{s}$
state. With this assignment, the $D_s \eta$ mode has a width of 13
MeV and becomes significant. It will helpful to search for this
mode experimentally.

If $D_{sJ}(2860)$ is the $1^{-}(2^{3}S_{1})$ candidate, the
dominant decay modes are $DK^\ast$ and $D^\ast K$. The decay width
of the DK mode is only 12 keV and tiny. Moreover its mass is much
higher than the quark model prediction $(2658\pm 15)$ MeV
\cite{zzx}. Its total width is around 90 MeV, significantly larger
than the experimental value $\Gamma=(48\pm 7\pm10)$ MeV.
Similarly, the total width becomes 132 MeV if $D_{sJ}(2860)$ is
the $1^{-}(1^{3}D_{1})$ candidate. The ratio of its two-body decay
modes is
$\Gamma(DK):\Gamma(D_{s}\eta):\Gamma(D^{*}K):\Gamma(D^{*}_{s}\eta):\Gamma(DK^{*})\approx
22:1:13:0.3:0.7$. We tend to conclude these two assignments are
not favorable for $D_{sJ}(2860)$ based on the available
experimental information.

If $D_{sJ}(2860)$ is the $0^{+}(2^{3}P_{0})$ $c\bar{s}$ state, its
total width is around 54 MeV, roughly consistent with the
experimental value considering theoretical uncertainty. The
dominant mode is $DK$ with a width of 37 MeV. The $D_s\eta$ mode
also has a very large width of 16 MeV. The $D_s^\ast\pi\pi$
three-body mode has a quite large width of 1.6 MeV. At present
only the $DK$ decay mode is observed experimentally.

If $D_{sJ}(2860)$ is the $3^{-}(1^{3}D_{3})$ $c\bar{s}$ state, its
total width is 37 MeV, also compatible with the experimental data.
Its dominant decay mode is $DK$. But now $D^\ast K$ mode has a
large width of 13 MeV. The ratio of its two-body decay modes is
$\Gamma(DK):\Gamma(D_{s}\eta):\Gamma(D^{*}K):\Gamma(D^{*}_{s}\eta):\Gamma(DK^{*})\approx
22:1:13:0.3:0.7$.

Naively one would expect both $0^{+}(2^{3}P_{0})$ and
$3^{-}(1^{3}D_{3})$ $c\bar{s}$ states to lie around 2860 MeV. The
available experimental information is not enough to distinguish
these possibilities. However, the $0^{+}(2^{3}P_{0})$ $c\bar{s}$
state does not decay into $DK^{*}$, $D^{*}K$ and $D_{s}^{*}\eta$.
The $3^{-}(1^{3}D_{3})$ $c\bar{s}$ state does not have a large
$D_s\eta$ decay width. Therefore, experimental search of
$D_{sJ}(2860)$ in the channels $D_s\eta$, $DK^{*}$, $D^{*}K$ and
$D_{s}^{*}\eta$ is strongly called for.

\section*{Acknowledgments}

This project was supported by the National Natural Science
Foundation of China under Grants 10625521 and 10421503, Key Grant
Project of Chinese Ministry of Education (No. 305001).

\section*{Appendix A: The $0^{+}(2^{3}P_{0})$ case}

\subsection{The expressions of $I^{0,0}_{0,0}$, $I^{1,-1}_{0,0}$,
$I^{-1,1}_{0,0}$, $I^{0,1}_{0,1}$, $I^{0,-1}_{0,-1}$,
$I^{1,0}_{0,1}$ and $I^{-1,0}_{0,-1}$ for $c\bar{s}(2^{3}P_{0})\to
D_{s}^{*}f_{0}(980)\to D_{s}^{*}\pi\pi$}
\begin{eqnarray}
I^{0,0}_{0,0}&=&-4\sqrt{\frac{6}{5}}\frac{|\mathbf{k}_{B}|\pi^{1/4}R_{A}^{5/2}R_{B}^{3/2}R_{C}^{5/2}}
{(R_{A}^{2}+R_{B}^{2}+R_{C}^{2})^{7/2}}\Bigg\{
R_{A}^{2}\Big[\frac{1}{16}(R_{A}^{2}+R_{B}^{2}+R_{C}^{2})^{2}(\xi-1)^{3}\xi(\xi+1)\mathbf{k}_{B}^{4}\nonumber\\&&+\frac{1}{2}
R_{A}^{2}+R_{B}^{2}+R_{C}^{2})(6\xi^{3}-6\xi^{2}-\xi+1)\mathbf{k}_{B}^{2}+21\xi-6\Big]-\frac{5}{8}(R_{A}^{2}+R_{B}^{2}+R_{C}^{2})
\xi\Big[(R_{A}^{2}+R_{B}^{2}+R_{C}^{2})\nonumber\\&&\times(\xi^{2}-1)\mathbf{k}_{B}^{2}+12\Big]\Bigg\}
\exp\bigg[-\frac{\mathbf{k}_{B}^{2}R_{A}^{2}(R_{B}^{2}+R_{C}^{2})}{8(R_{A}^{2}+R_{B}^{2}+R_{C}^{2})}\bigg],
\end{eqnarray}
\begin{eqnarray}
I^{1,-1}_{0,0}&=&I^{-1,1}_{0,0}=\sqrt{\frac{6}{5}}\frac{|\mathbf{k}_{B}|\pi^{1/4}R_{A}^{5/2}R_{B}^{3/2}R_{C}^{5/2}}
{(R_{A}^{2}+R_{B}^{2}+R_{C}^{2})^{7/2}}\Bigg\{ \mathbf{k}_{B}^2
(\xi-1)^{2}\xi
R_{A}^{4}+R_{A}^{2}\bigg[\Big(\mathbf{k}_{B}^{2}(R_{B}^{2}+R_{C}^{2})(\xi-1)^{2}+18\Big)\xi-8\bigg]
\nonumber\\&&-10(R_{B}^{2}+R_{C}^{2})\xi\Bigg\}
\exp\bigg[-\frac{\mathbf{k}_{B}^{2}R_{A}^{2}(R_{B}^{2}+R_{C}^{2})}{8(R_{A}^{2}+R_{B}^{2}+R_{C}^{2})}\bigg],
\end{eqnarray}
\begin{eqnarray}
I^{0,1}_{0,1}&=&I^{0,-1}_{0,-1}=-\sqrt{\frac{6}{5}}\frac{|\mathbf{k}_{B}|\pi^{1/4}R_{A}^{5/2}R_{B}^{3/2}R_{C}^{5/2}}
{(R_{A}^{2}+R_{B}^{2}+R_{C}^{2})^{7/2}}\Bigg\{
\Big[R_{A}^{2}\Big(\mathbf{k}_{B}^{2}(R_{A}^{2}+R_{B}^{2}+R_{C}^{2})(\xi-1)^{2}+18\Big)\nonumber\\&&
-10(R_{B}^{2}+R_{C}^{2})\Big]
(\xi-1)\Bigg\}\exp\bigg[-\frac{\mathbf{k}_{B}^{2}R_{A}^{2}(R_{B}^{2}+R_{C}^{2})}{8(R_{A}^{2}+R_{B}^{2}+R_{C}^{2})}\bigg],
\end{eqnarray}
\begin{eqnarray}
I^{1,0}_{0,1}&=&I^{-1,0}_{0,-1}=-4\sqrt{\frac{6}{5}}\frac{|\mathbf{k}_{B}|\pi^{1/4}R_{A}^{5/2}R_{B}^{3/2}R_{C}^{5/2}}
{(R_{A}^{2}+R_{B}^{2}+R_{C}^{2})^{7/2}}\Bigg\{
R_{A}^{2}\Big[\frac{1}{4}\mathbf{k}_{B}^{2}(R_{A}^{2}+R_{B}^{2}+R_{C}^{2})(\xi+1)(\xi-1)^{2}+7\xi+3
\Big]\nonumber\\&&
-\frac{5}{2}(R_{A}^{2}+R_{B}^{2}+R_{C}^{2})(\xi+1)\Bigg\}
\exp\bigg[-\frac{\mathbf{k}_{B}^{2}R_{A}^{2}(R_{B}^{2}+R_{C}^{2})}{8(R_{A}^{2}+R_{B}^{2}+R_{C}^{2})}\bigg]
\end{eqnarray}
with $\xi=\frac{R_{A}^{2}}{R_{A}^{2}+R_{B}^{2}+R_{C}^{2}}$, where
the indexes $A$, $B$ and $C$ correspond to $D_{sJ}(2860)$,
$D^{*}_{s}$ and $f_{0}(980)$.

\subsection{ The expressions of $I^{M_{L_A},m}_{M_{L_B},M_{L_C}}$
appeared in the calculation of $c\bar{s}(2^{3}P_{0})\to
D_{sJ}^{*}(2317)f_{0}(980)\to D_{sJ}^{*}(2317)\pi\pi$}
\begin{eqnarray}
I^{0,0}_{0,0}&=&-16\sqrt{\frac{3}{5}}\frac{\pi^{1/4}R_{A}^{5/2}R_{B}^{5/2}R_{C}^{5/2}}
{(R_{A}^{2}+R_{B}^{2}+R_{C}^{2})^{9/2}}\Bigg\{
\frac{1}{64}R_{A}^{2}\Big[(R_{A}^{2}+R_{B}^{2}+R_{C}^{2})^{3}(\xi-1)^{3}\xi^{2}(\xi+1)\mathbf{k}_{B}^{6}\nonumber\\&&
+4(R_{A}^{2}+R_{B}^{2}+R_{C}^{2})^{2}(17\xi^{4}-20\xi^{3}-2\xi^{2}+6\xi-1)\mathbf{k}_{B}^{4}
+16(R_{A}^{2}+R_{B}^{2}+R_{C}^{2})(57\xi^{2}-30\xi-2)\mathbf{k}_{B}^{2}+1344
\Big]
\nonumber\\&&-\frac{5}{2}(R_{A}^{2}+R_{B}^{2}+R_{C}^{2})\Big[
\frac{1}{16}(R_{A}^{2}+R_{B}^{2}+R_{C}^{2})^{2}\xi^{2}(\xi^{2}-1)\mathbf{k}_{B}^{4}
+\frac{1}{4}(R_{A}^{2}+R_{B}^{2}+R_{C}^{2})(6\xi^{2}-1)\mathbf{k}_{B}^{2}+3
\Big]\Bigg\}\nonumber\\&&\times
\exp\bigg[-\frac{\mathbf{k}_{B}^{2}R_{A}^{2}(R_{B}^{2}+R_{C}^{2})}{8(R_{A}^{2}+R_{B}^{2}+R_{C}^{2})}\bigg],
\end{eqnarray}
\begin{eqnarray}
I^{0,1}_{0,1}&=&I^{0,1}_{1,0}=I^{0,-1}_{0,-1}=I^{0,-1}_{-1,0}
\nonumber\\&&=-16\sqrt{\frac{3}{5}}\frac{\pi^{1/4}R_{A}^{5/2}R_{B}^{5/2}R_{C}^{5/2}}
{(R_{A}^{2}+R_{B}^{2}+R_{C}^{2})^{9/2}}\Bigg\{ R_{A}^{2}\Big[
\frac{1}{16}(R_{A}^{2}+R_{B}^{2}+R_{C}^{2})\Big(\xi[\mathbf{k}_{B}^{2}(R_{A}^{2}+R_{B}^{2}+R_{C}^{2})(\xi-1)^{3}+40\xi
-52]\nonumber\\&&+12\Big)\mathbf{k}_{B}^{2}+7\Big]-\frac{5}{8}(R_{A}^{2}+R_{B}^{2}+R_{C}^{2})
\bigg[(R_{A}^{2}+R_{B}^{2}+R_{C}^{2})(\xi-1)\xi\mathbf{k}_{B}^{2}+4\bigg]\Bigg\}
\exp\bigg[-\frac{\mathbf{k}_{B}^{2}R_{A}^{2}(R_{B}^{2}+R_{C}^{2})}{8(R_{A}^{2}+R_{B}^{2}+R_{C}^{2})}\bigg],
\end{eqnarray}
\begin{eqnarray}
I^{1,0}_{0,1}&=&I^{1,0}_{1,0}=I^{-1,0}_{0,-1}=I^{-1,0}_{-1,0}
\nonumber\\&&=-16\sqrt{\frac{3}{5}}\frac{\pi^{1/4}R_{A}^{5/2}R_{B}^{5/2}R_{C}^{5/2}}
{(R_{A}^{2}+R_{B}^{2}+R_{C}^{2})^{9/2}}\Bigg\{ R_{A}^{2}\Big[
\frac{1}{16}(R_{A}^{2}+R_{B}^{2}+R_{C}^{2})\Big(\xi[\mathbf{k}_{B}^{2}(R_{A}^{2}+R_{B}^{2}+R_{C}^{2})(\xi+1)
(\xi-1)^{2}+40\xi
+4]\nonumber\\&&-4\Big)\mathbf{k}_{B}^{2}+7\Big]-\frac{5}{8}(R_{A}^{2}+R_{B}^{2}+R_{C}^{2})
\bigg[(R_{A}^{2}+R_{B}^{2}+R_{C}^{2})(\xi+1)\xi\mathbf{k}_{B}^{2}+4\bigg]\Bigg\}
\exp\bigg[-\frac{\mathbf{k}_{B}^{2}R_{A}^{2}(R_{B}^{2}+R_{C}^{2})}{8(R_{A}^{2}+R_{B}^{2}+R_{C}^{2})}\bigg],
\end{eqnarray}
\begin{eqnarray}
I^{1,-1}_{0,0}&=&I^{-1,1}_{0,0}\nonumber\\
&=&16\sqrt{\frac{3}{5}}\frac{\pi^{1/4}R_{A}^{5/2}R_{B}^{5/2}R_{C}^{5/2}}
{(R_{A}^{2}+R_{B}^{2}+R_{C}^{2})^{9/2}}\Bigg\{ R_{A}^{2}\Big[
\frac{1}{16}(R_{A}^{2}+R_{B}^{2}+R_{C}^{2})\Big(\xi
\big([\mathbf{k}_{B}^{2}(R_{A}^{2}+R_{B}^{2}+R_{C}^{2})
(\xi-1)^{2}+40]\xi-24\big)+4\Big)\mathbf{k}_{B}^{2}\nonumber\\&&+7\Big]-\frac{5}{8}(R_{A}^{2}+R_{B}^{2}+R_{C}^{2})
\bigg[(R_{A}^{2}+R_{B}^{2}+R_{C}^{2})\xi^{2}\mathbf{k}_{B}^{2}+4\bigg]\Bigg\}
\exp\bigg[-\frac{\mathbf{k}_{B}^{2}R_{A}^{2}(R_{B}^{2}+R_{C}^{2})}{8(R_{A}^{2}+R_{B}^{2}+R_{C}^{2})}\bigg],
\end{eqnarray}
\begin{eqnarray}
I^{0,0}_{1,-1}&=&I^{0,0}_{-1,1}\nonumber\\&=&
16\sqrt{\frac{3}{5}}\frac{\pi^{1/4}R_{A}^{5/2}R_{B}^{5/2}R_{C}^{5/2}}
{(R_{A}^{2}+R_{B}^{2}+R_{C}^{2})^{9/2}}\Bigg\{ R_{A}^{2}\Big[
\frac{1}{16}(R_{A}^{2}+R_{B}^{2}+R_{C}^{2})\Big(
\mathbf{k}_{B}^{2} (R_{A}^{2}+R_{B}^{2}+R_{C}^{2})(\xi+1)(\xi
-1)^{3}\nonumber\\&& +8(5\xi+2)(\xi-1)\Big)\mathbf{k}_{B}^{2}+7
\Big] -\frac{5}{8}(R_{A}^{2}+R_{B}^{2}+R_{C}^{2})
\bigg[(R_{A}^{2}+R_{B}^{2}+R_{C}^{2})(\xi^{2}-1)\mathbf{k}_{B}^{2}+4
\bigg]\Bigg\}
\nonumber\\&&\times\exp\bigg[-\frac{\mathbf{k}_{B}^{2}R_{A}^{2}(R_{B}^{2}+R_{C}^{2})}{8(R_{A}^{2}
+R_{B}^{2}+R_{C}^{2})}\bigg],
\end{eqnarray}
\begin{eqnarray}
I^{1,1}_{1,1}&=&I^{-1,-1}_{-1,-1}=
-8\sqrt{\frac{3}{5}}\frac{\pi^{1/4}R_{A}^{5/2}R_{B}^{5/2}R_{C}^{5/2}}
{(R_{A}^{2}+R_{B}^{2}+R_{C}^{2})^{9/2}}\Bigg\{
R_{A}^{2}\Big[\mathbf{k}_{B}^{2}
(R_{A}^{2}+R_{B}^{2}+R_{C}^{2})(\xi-1)^{2}+18\Big]-10(R_{B}^{2}+R_{C}^{2})\Bigg\}.
\end{eqnarray}

\section*{Appendix B: The $1^{-}(1^{3}D_{1})$ case}
\subsection{The relevant expressions in the calculation of
$c\bar{s}(1^{3}D_{1})\to D_{sJ}^{*}(2317)f_{0}(980)\to
D_{sJ}^{*}(2317)\pi\pi$}
\begin{eqnarray}
I^{-2,0}_{-1,-1}=I^{2,0}_{1,1}=-8\sqrt{6}\frac{|\mathbf{k}_{B}|\pi^{1/4}R_{A}^{7/2}R_{B}^{5/2}R_{C}^{5/2}}
{(R_{A}^{2}+R_{B}^{2}+R_{C}^{2})^{7/2}} (1+\xi)
\exp\bigg[-\frac{\mathbf{k}_{B}^{2}R_{A}^{2}(R_{B}^{2}+R_{C}^{2})}{8(R_{A}^{2}+R_{B}^{2}+R_{C}^{2})}\bigg],
\end{eqnarray}
\begin{eqnarray}
I^{-2,1}_{-1,0}=I^{-2,1}_{0,-1}=I^{2,-1}_{0,1}=I^{2,-1}_{1,0}=8\sqrt{6}\frac{|\mathbf{k}_{B}|\pi^{1/4}R_{A}^{7/2}R_{B}^{5/2}R_{C}^{5/2}}
{(R_{A}^{2}+R_{B}^{2}+R_{C}^{2})^{7/2}} \xi
\exp\bigg[-\frac{\mathbf{k}_{B}^{2}R_{A}^{2}(R_{B}^{2}+R_{C}^{2})}{8(R_{A}^{2}+R_{B}^{2}+R_{C}^{2})}\bigg],
\end{eqnarray}
\begin{eqnarray}
I^{-1,-1}_{-1,-1}&=&I^{-1,1}_{-1,1}=I^{-1,1}_{1,-1}=I^{1,-1}_{-1,1}=I^{1,-1}_{1,-1}=I^{1,1}_{1,1}=-16\sqrt{3}\frac{|\mathbf{k}_{B}|\pi^{1/4}R_{A}^{7/2}R_{B}^{5/2}R_{C}^{5/2}}
{(R_{A}^{2}+R_{B}^{2}+R_{C}^{2})^{7/2}} (-1+\xi)
\nonumber\\&&\times
\exp\bigg[-\frac{\mathbf{k}_{B}^{2}R_{A}^{2}(R_{B}^{2}+R_{C}^{2})}{8(R_{A}^{2}+R_{B}^{2}+R_{C}^{2})}\bigg],
\end{eqnarray}
\begin{eqnarray}
I^{-1,0}_{-1,0}&=&I^{-1,0}_{0,-1}=I^{1,0}_{0,1}=I^{1,0}_{1,0}=-2\sqrt{3}\frac{|\mathbf{k}_{B}|\pi^{1/4}R_{A}^{7/2}R_{B}^{5/2}R_{C}^{5/2}}
{(R_{A}^{2}+R_{B}^{2}+R_{C}^{2})^{7/2}} \xi \Bigg\{
12+\mathbf{k}_{B}^2(R_{A}^{2}+R_{B}^{2}+R_{C}^{2})(-1+\xi^2) \Bigg\}
\nonumber\\&&\times\exp\bigg[-\frac{\mathbf{k}_{B}^{2}R_{A}^{2}(R_{B}^{2}+R_{C}^{2})}{8(R_{A}^{2}+R_{B}^{2}+R_{C}^{2})}\bigg],
\end{eqnarray}
\begin{eqnarray}
I^{-1,1}_{0,0}=I^{1,-1}_{0,0}=2\sqrt{3}\frac{|\mathbf{k}_{B}|\pi^{1/4}R_{A}^{7/2}R_{B}^{5/2}R_{C}^{5/2}}
{(R_{A}^{2}+R_{B}^{2}+R_{C}^{2})^{7/2}} \Bigg\{
-4+12\xi+\mathbf{k}_{B}^2 R_A^2 \xi (-1+\xi) \Bigg\}
\exp\bigg[-\frac{\mathbf{k}_{B}^{2}R_{A}^{2}(R_{B}^{2}+R_{C}^{2})}{8(R_{A}^{2}+R_{B}^{2}+R_{C}^{2})}\bigg],
\end{eqnarray}

\begin{eqnarray}
I^{0,-1}_{-1,0}=I^{0,-1}_{0,-1}=I^{0,1}_{0,1}=I^{0,1}_{1,0}=-2
\frac{|\mathbf{k}_{B}|\pi^{1/4}R_{A}^{7/2}R_{B}^{5/2}R_{C}^{5/2}}
{(R_{A}^{2}+R_{B}^{2}+R_{C}^{2})^{7/2}} \Bigg\{
-8+4\xi+\mathbf{k}_{B}^2 R_A^2 (-1+\xi)^2 \Bigg\}
\exp\bigg[-\frac{\mathbf{k}_{B}^{2}R_{A}^{2}(R_{B}^{2}+R_{C}^{2})}{8(R_{A}^{2}+R_{B}^{2}+R_{C}^{2})}\bigg],
\end{eqnarray}

\begin{eqnarray}
I^{0,0}_{-1,1}&=&I^{0,0}_{1,-1}=2
\frac{|\mathbf{k}_{B}|\pi^{1/4}R_{A}^{7/2}R_{B}^{5/2}R_{C}^{5/2}}
{(R_{A}^{2}+R_{B}^{2}+R_{C}^{2})^{7/2}} \Bigg\{ 4(-3+\xi)+
\mathbf{k}_{B}^2 (R_A^2+R_B^2+R_C^2)(-1+\xi)^2 (1+\xi) \Bigg\}
\nonumber\\&&\times
\exp\bigg[-\frac{\mathbf{k}_{B}^{2}R_{A}^{2}(R_{B}^{2}+R_{C}^{2})}{8(R_{A}^{2}+R_{B}^{2}+R_{C}^{2})}\bigg],
\end{eqnarray}

\begin{eqnarray}
I^{0,0}_{0,0}&=&-
\frac{|\mathbf{k}_{B}|\pi^{1/4}R_{A}^{7/2}R_{B}^{5/2}R_{C}^{5/2}}
{2(R_{A}^{2}+R_{B}^{2}+R_{C}^{2})^{7/2}} \Bigg\{
64(-1+3\xi)+\mathbf{k}_{B}^2
(R_A^2+R_B^2+R_C^2)(-1+\xi)\Big[-4+\mathbf{k}_{B}^2\frac{R_A^4(-1+\xi^2)}{R_A^2+R_B^2+R_C^2}+4\xi(2+9\xi)\Big]
\Bigg\}\nonumber\\&&\times
\exp\bigg[-\frac{\mathbf{k}_{B}^{2}R_{A}^{2}(R_{B}^{2}+R_{C}^{2})}{8(R_{A}^{2}+R_{B}^{2}+R_{C}^{2})}\bigg],
\end{eqnarray}

\subsection{The relevant expressions in the calculation of
$c\bar{s}(1^{3}D_{1})\to D_{s}^{*}f_{0}(980)\to D_{s}^{*}\pi\pi$}
\begin{eqnarray}
I^{-2,1}_{0,-1}=I^{2,-1}_{0,1}=16\sqrt{3}\frac{\pi^{1/4}R_{A}^{7/2}R_{B}^{3/2}R_{C}^{5/2}}{{(R_{A}^{2}+R_{B}^{2}+R_{C}^{2})^{7/2}}}
\exp\bigg[-\frac{\mathbf{k}_{B}^{2}R_{A}^{2}(R_{B}^{2}+R_{C}^{2})}{8(R_{A}^{2}+R_{B}^{2}+R_{C}^{2})}\bigg],
\end{eqnarray}
\begin{eqnarray}
I^{-1,0}_{0,-1}=I^{1,0}_{0,1}=-2\sqrt{6}\frac{\pi^{1/4}R_{A}^{7/2}R_{B}^{3/2}R_{C}^{5/2}}{{(R_{A}^{2}+R_{B}^{2}+R_{C}^{2})^{7/2}}}
\bigg[  4+  \mathbf{k}_{B}^{2}  (R_{A}^{2}+R_{B}^{2}+R_{C}^{2})
(-1+\xi^2) \bigg]
\exp\bigg[-\frac{\mathbf{k}_{B}^{2}R_{A}^{2}(R_{B}^{2}+R_{C}^{2})}{8(R_{A}^{2}+R_{B}^{2}+R_{C}^{2})}\bigg],
\end{eqnarray}
\begin{eqnarray}
I^{-1,1}_{0,0}=I^{1,-1}_{0,0}=2\sqrt{6}\frac{\pi^{1/4}R_{A}^{7/2}R_{B}^{3/2}R_{C}^{5/2}}{{(R_{A}^{2}+R_{B}^{2}+R_{C}^{2})^{7/2}}}
\bigg[  4+  \mathbf{k}_{B}^{2}  R_{A}^{2} (-1+\xi) \bigg]
\exp\bigg[-\frac{\mathbf{k}_{B}^{2}R_{A}^{2}(R_{B}^{2}+R_{C}^{2})}{8(R_{A}^{2}+R_{B}^{2}+R_{C}^{2})}\bigg],
\end{eqnarray}
\begin{eqnarray}
I^{0,-1}_{0,-1}=I^{0,1}_{0,1}=-2\sqrt{2}\frac{\pi^{1/4}R_{A}^{7/2}R_{B}^{3/2}R_{C}^{5/2}}{{(R_{A}^{2}+R_{B}^{2}+R_{C}^{2})^{7/2}}}
\bigg[  -4+  \mathbf{k}_{B}^{2}  (R_{A}^{2}+R_{B}^{2}+R_{C}^{2})
(-1+\xi)^2 \bigg]
\exp\bigg[-\frac{\mathbf{k}_{B}^{2}R_{A}^{2}(R_{B}^{2}+R_{C}^{2})}{8(R_{A}^{2}+R_{B}^{2}+R_{C}^{2})}\bigg],
\end{eqnarray}
\begin{eqnarray}
I^{0,0}_{0,0}&=&-\frac{1}{\sqrt{2}}
    \frac{\pi^{1/4}R_{A}^{7/2}R_{B}^{3/2}R_{C}^{5/2}}{{(R_{A}^{2}+R_{B}^{2}+R_{C}^{2})^{7/2}}}
\bigg\{  32 +  \mathbf{k}_{B}^{2}  (R_{A}^{2}+R_{B}^{2}+R_{C}^{2})
(-1+\xi) [4+ 20 \xi + \mathbf{k}_{B}^{2} R_A^2 (-1+ \xi^2)] \bigg\}
\nonumber\\&&\times\exp\bigg[-\frac{\mathbf{k}_{B}^{2}R_{A}^{2}(R_{B}^{2}+R_{C}^{2})}{8(R_{A}^{2}+R_{B}^{2}+R_{C}^{2})}\bigg],
\end{eqnarray}

\section*{Appendix C: The $1^{-}(2^{3}S_{1})$ case}

\subsection{The expressions of $I^{M_{L_A},m}_{M_{L_B},M_{L_C}}$
in Eq. (\ref{expression-4})}
\begin{eqnarray}
I^{0,-1}_{-1,0}&=& I^{0,-1}_{0,-1}=  I^{0,1}_{0,1}= I^{0,1}_{1,0}=
\sqrt{2} \frac{|\mathbf{k}_{B}|
    \pi^{1/4}R_{A}^{3/2}R_{B}^{5/2}R_{C}^{5/2}}
{(R_{A}^{2}+R_{B}^{2}+R_{C}^{2})^{7/2}}             \Bigg\{
-6R_A^2+R_A^2 \bigg[  -8+ 28\xi+ \mathbf{k}_{B}^2 R_A^2 (-1+\xi)^2
\bigg]
 \Bigg\}\nonumber\\&&\times
\exp\bigg[-\frac{\mathbf{k}_{B}^{2}R_{A}^{2}(R_{B}^{2}+R_{C}^{2})}{8(R_{A}^{2}+R_{B}^{2}+R_{C}^{2})}\bigg],
\end{eqnarray}
\begin{eqnarray}
I^{0,0}_{-1,1}&=& I^{0,0}_{1,-1} - \sqrt{2} \frac{|\mathbf{k}_{B}|
    \pi^{1/4} R_{A}^{3/2}R_{B}^{5/2}R_{C}^{5/2}}
{(R_{A}^{2}+R_{B}^{2}+R_{C}^{2})^{7/2}}             \Bigg\{
-6(R_{A}^{2}+R_{B}^{2}+R_{C}^{2})(1+\xi)+R_A^2 \bigg[  12+
28\xi\nonumber\\&&+ \mathbf{k}_{B}^2
(R_{A}^{2}+R_{B}^{2}+R_{C}^{2}) (-1+\xi)^2 (1+\xi) \bigg]
 \Bigg\}
\exp\bigg[-\frac{\mathbf{k}_{B}^{2}R_{A}^{2}(R_{B}^{2}+R_{C}^{2})}{8(R_{A}^{2}+R_{B}^{2}+R_{C}^{2})}\bigg],
\end{eqnarray}
\begin{eqnarray}
I^{0,0}_{0,0}&=&  \frac{1}{2\sqrt{2}} \frac{|\mathbf{k}_{B}|
    \pi^{1/4} R_{A}^{3/2}R_{B}^{5/2}R_{C}^{5/2}}
{(R_{A}^{2}+R_{B}^{2}+R_{C}^{2})^{7/2}}             \Bigg\{
-6\mathbf{k}_{B}^2 R_A^4(1+\xi)
-24(R_{A}^{2}+R_{B}^{2}+R_{C}^{2})(1+3\xi)\nonumber\\&&+R_A^2[-16+336\xi+\mathbf{k}_{B}^4
R_A^4(-1+\xi)^2(1+\xi)+4\mathbf{k}_{B}^2(R_{A}^{2}+R_{B}^{2}+R_{C}^{2})(-1+2\xi)(1+2\xi)(-1+3\xi)]
 \Bigg\}\nonumber\\&&\times
\exp\bigg[-\frac{\mathbf{k}_{B}^{2}R_{A}^{2}(R_{B}^{2}+R_{C}^{2})}{8(R_{A}^{2}+R_{B}^{2}+R_{C}^{2})}\bigg],
\end{eqnarray}

\subsection{The expressions of $I^{M_{L_A},m}_{M_{L_B},M_{L_C}}$
in Eq. (\ref{expression-5})}
\begin{eqnarray}
I^{0,-1}_{0,-1}&=& I^{0,1}_{0,1}= 2 \frac{
    \pi^{1/4} R_{A}^{3/2}R_{B}^{3/2}R_{C}^{5/2}}
{(R_{A}^{2}+R_{B}^{2}+R_{C}^{2})^{7/2}}             \Bigg\{
-6(R_{A}^{2}+R_{B}^{2}+R_{C}^{2})+R_A^2 \bigg[  20+
\mathbf{k}_{B}^{2}(R_{A}^{2}+R_{B}^{2}+R_{C}^{2})(-1+\xi)^2 \bigg]
 \Bigg\}\nonumber\\&&\times
\exp\bigg[-\frac{\mathbf{k}_{B}^{2}R_{A}^{2}(R_{B}^{2}+R_{C}^{2})}{8(R_{A}^{2}+R_{B}^{2}+R_{C}^{2})}\bigg],
\end{eqnarray}
\begin{eqnarray}
I^{0,0}_{0,0}&=&  \frac{1}{2} \frac{
    \pi^{1/4} R_{A}^{3/2}R_{B}^{3/2}R_{C}^{5/2}}
{(R_{A}^{2}+R_{B}^{2}+R_{C}^{2})^{7/2}}             \Bigg\{
80R_A^2-24(R_{A}^{2}+R_{B}^{2}+R_{C}^{2})+\mathbf{k}_{B}^{2}(R_{A}^{2}+R_{B}^{2}+R_{C}^{2})
\bigg[-6R_A^2(1+\xi)\nonumber\\&&+R_A^2 \bigg( -4+R_A^2
\mathbf{k}_{B}^{2} (-1+\xi)^2(1+\xi)+4\xi(-1+8\xi) \bigg) \bigg]
 \Bigg\}
\exp\bigg[-\frac{\mathbf{k}_{B}^{2}R_{A}^{2}(R_{B}^{2}+R_{C}^{2})}{8(R_{A}^{2}+R_{B}^{2}+R_{C}^{2})}\bigg].
\end{eqnarray}

\end{document}